\documentclass[fleqn,usenatbib]{mnras}

\usepackage[T1]{fontenc}
\usepackage{ae,aecompl}

\usepackage{graphicx}	% Including figure files
\usepackage{amsmath}	% Advanced maths commands
\usepackage{amssymb}	% Extra maths symbols
\usepackage{xspace}
\usepackage{calc}

\PassOptionsToPackage{hyphens}{url}
\usepackage{hyperref}
\usepackage{url}

\newcommand{\ac}{Astron. \& Comput.\xspace}

\newcommand{\Var}{\mathop{\text{Var}}\nolimits}
\title[WASP-33 photometric variability]{Cleaning WASP-33\,\emph{b} transits from the host
star photometric variability: analysis of TESS data from two sectors}

\author[R.V.~Baluev \& E.N.~Sokov]{%
Roman~V. Baluev$^{1,2}$\thanks{E-mail: r.baluev@spbu.ru} and Eugene~N. Sokov$^3$\\
$^1$Saint Petersburg State University, 7--9 Universitetskaya nab., St Petersburg 199034, Russia\\
$^2$Special Astrophysical Observatory, Russian Academy of Sciences, Nizhnii Arkhyz, 369167, Russia\\
$^3$Central Astronomical Observatory at Pulkovo of Russian Academy of Sciences, Pulkovskoje sh. 65/1, St Petersburg 196140, Russia}

\begin{document}

\date{Accepted 2024 December 31.
      Received 2024 December 30;
      in original form 2024 September 15}

\pagerange{\pageref{firstpage}--\pageref{lastpage}} \pubyear{2024}

\maketitle

\label{firstpage}

\begin{abstract}
Based on TESS observations of a $\delta$~Scuti variable WASP-33 obtained in 2019 and 2022,
we thoroughly investigate the power spectrum of this target photometric flux and construct
a statistically exhaustive model for its variability. This model contains $30$ robustly
justified harmonics detected in the both TESS sectors simultaneously, $13$ less robust
harmonics detected in a single TESS sector, a red noise and a quasiperiodic noise terms.
This allowed us to greatly improve the accuracy of the exoplanet WASP-33\,\emph{b} transit
timings, reducing the TTV residuals r.m.s. drastically, by a factor of $3.5$, from $63$~s
to $18$~s. Finally, our analysis does not confirm existence of a detectable orbital phase
variation claimed previously based on WASP-33 TESS photometry of 2019.
\end{abstract}


\section{Introduction}
WASP-33 (HD~15082) is a bright 8.14 mag fast-rotating star with $v \sin i \sim 90$~km/s.
Its only known exoplanet WASP-33\,\emph{b} was detected by SuperWASP \citep{Herrero11}
with orbital period currently estimated by $P_b = (1.219867\pm 4.2\cdot 10^{-5})$~d
\citep{vonEssen20}, mass by $M_b = (2.10\pm 0.14)$~$M_{\rm Jup}$, and radius by $r_b =
(1.59\pm 0.07)$~$R_{\rm Jup}$ \citep{ChakrabartySengupta19}. It is remarkable that this
exoplanet was the first one ever known to orbit a $\delta$~Scuti variable.

The variability of the prototype star itself, $\delta$~Scuti, was discovered in the
beginning of the XX century \citep{CampbellWright1900}, later confirmed by spectral and
photometric observations \citep{Colacevich35,Fath35}. Furthermore, \citet{Eggen56} combined
this and a few similar stars together in a standalone class of variables, and so the
$\delta$~Scuti type of variable stars appeared.

These variables are pulsating stars belonging to spectral classes A0-F5 III-IV, located in
the instability strip of the HR diagram \citep{Baglin73,BregerStockenhuber83}. Their $m_V$
variation amplitudes range from $0.003$ to $0.9$ (typically a few hundreds) and periods
from $0.01$~d to $0.2$~d. Most of the $\delta$~Scuti stars belong to type I population
\citep{Breger79}, with a typical mass of $\sim 2$~$M_\odot$ \citep{MilliganCarson92}. The
power spectrum of $\delta$~Scuti variables may reveal tens of periods \citep{Breger99}
ranging from hours to minutes. Such stars often exhibit significant changes in the form of
their lightcurve, in their variability period and amplitude (Blazhko effect). A key role in
the pulsation generating mechanism belongs to helium which is subject to heating/ionization
and cooling/recombination processes, thus changing its opacity in a cycle \citep{Cox63}.
Radial as well as non-radial pulsations can be observed.

Large amount of data regarding $\delta$~Scuti variables were brought by (i) OGLE and MACHO
surveys which detected about $15000$ such variables in the Large Magellanic Cloud
\citep{Poleski10,Garg10,Soszynski23}, (ii) Kepler/K2 missions \citep{Borucki10,Gusik21}
that found about $2000$ variables of this type, and (iii) TESS (Transiting Exoplanet Survey
Satellite) mission, for example \citet{Gootkin24} identified a sample of $\sim 14000$
variables likely belonging to the $\delta$~Scuti class.

The intrinsic variability of WASP-33 represents an obvious nuisance factor for analysing
its exoplanet \emph{b} transit lightcurves. After the discovery work by \citet{Herrero11},
the effect of this variability on the estimated planetary orbital parameters was
investigated using out-of-transit observations from ground-based telescopes located in
Germany and Spain \citep{vonEssen14}. Their investigation of the WASP-33 pulsations power
spectrum revealed eight significant frequencies. It was concluded that cleaning the transit
lightcurves from stellar pulsations practically does not affect estimated planetary
parameters, but their uncertainties get decreased. Besides, \citet{vonEssen14} noticed that
WASP-33 pulsations phases change over time. In a later analysis of TESS photometry of 2019,
\citet{vonEssen20} claimed the detection of $29$ pulsation frequencies. In addition to the
power spectrum analysis, they also reported statistically significant detection of
WASP-33\,\emph{b} secondary eclipse and of the orbital phase curve variation (period of
$P_b$, amplitude of $100.4 \pm 13.1$~ppm). Also, they took into account the ``ellipsoidal
variation'' that appears because of planetary gravitational effect on the star shape
(period of $P_b/2$, theoretically predicted amplitude of $27.4$~ppm).

A more recent analysis of WASP-33 variability was performed using data from TESS
\citep{Kalman22}. They investigated the Fourier spectrum of stellar oscillations
reconstructed with TLCM (Transit and Light Curve Modeller, \citealt{Csizmadia23}). This
spectrum revealed three peaks with $S/N>8$ that were close to a commensurability with the
planetary orbital period ($1:3$, $1:12$, and $1:25$). It was argued that the distribution
of $17$ frequencies with the largest amplitudes tend to shift right from these
subharmonics. As \citet{Kalman22} noticed, this effect can be caused by the tidal
perturbations from the planet.

Our primary goal here is to analyse WASP-33 pulsations spectrum based on the up-to-date
TESS data (observations of 2019 and 2022), and to investigate whether (and how much) this
can help us to improve the transit timings accuracy for the exoplanet WASP-33\,\emph{b}.
The paper contains three major sections, discussing subsequently the TESS observations in
Sect.~\ref{sec_obs}, construction of the photometric variation model in
Sect.~\ref{sec_analysis}, and applying this model to improve the set of estimated planet
parameters and transit timings, in Sect.~\ref{sec_transits}.

\section{TESS photometry of WASP-33}
\label{sec_obs}
\subsection{Basic setting}
\label{sec_dover}
TESS observed WASP-33 in sector~18 (Nov 2019) and sector~58 (Nov 2022), lasting about $1$
month each. In the DVT data downloaded from the MAST (Mikulski Archive for Space
Telescopes) we found $14796$ and $18951$ photometric measurements for these two sectors,
respectively. We identified $17$ transits in sector~18 and $21$ transits in sector~58 (plus
a partial transit in the sector end). For each transit we extracted a $\pm 4$~h piece of
the lightcurve about the predicted midtransit epoch. This subset formed our ``in-transit''
dataset (ITD). The ``out-of-transit'' dataset (OTD) included all observations located at
least $\pm 2$~h away from the centres of transits and secondary eclipses. Notice that ITD
and OTD slightly overlap.

In this work we will deal mainly with the OTD, since we aim to construct a model for the
intrinsic star variability. The sector~18 portion of the OTD will be hereafter refered to
as TESS$_{18}$, and it contains $10684$ datapoints. The sector~58 OTD portion (TESS$_{58}$)
contains $13625$ datapoints. ITD contains $8914$ datapoints in total, with $\sim 200$
observations per transit.

\begin{figure*}
\includegraphics[trim={0.45cm 0.35cm 0.55cm 0.35cm},clip,height=0.632\linewidth]{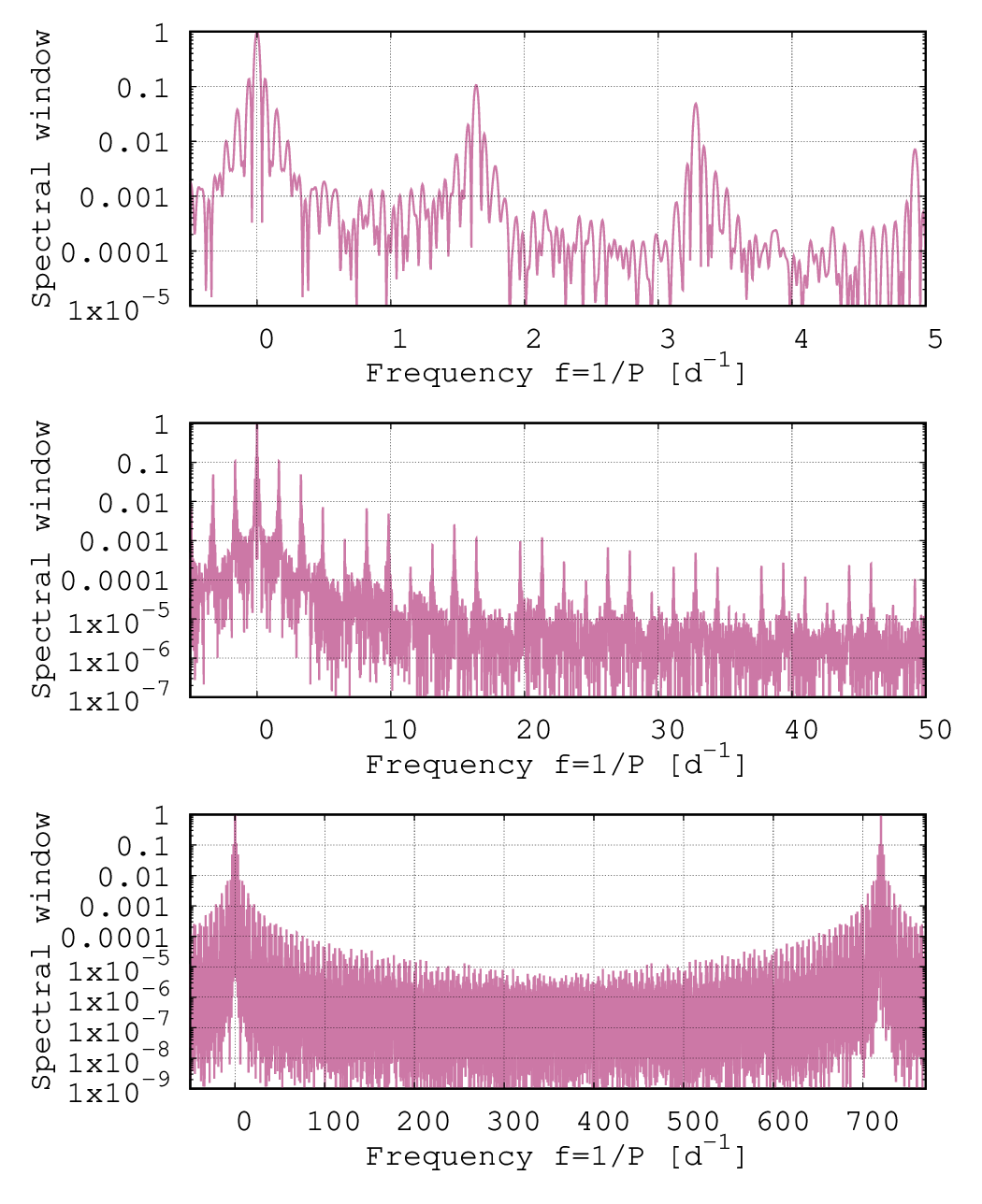}
\includegraphics[trim={1.05cm 0.35cm 0.55cm 0.35cm},clip,height=0.632\linewidth]{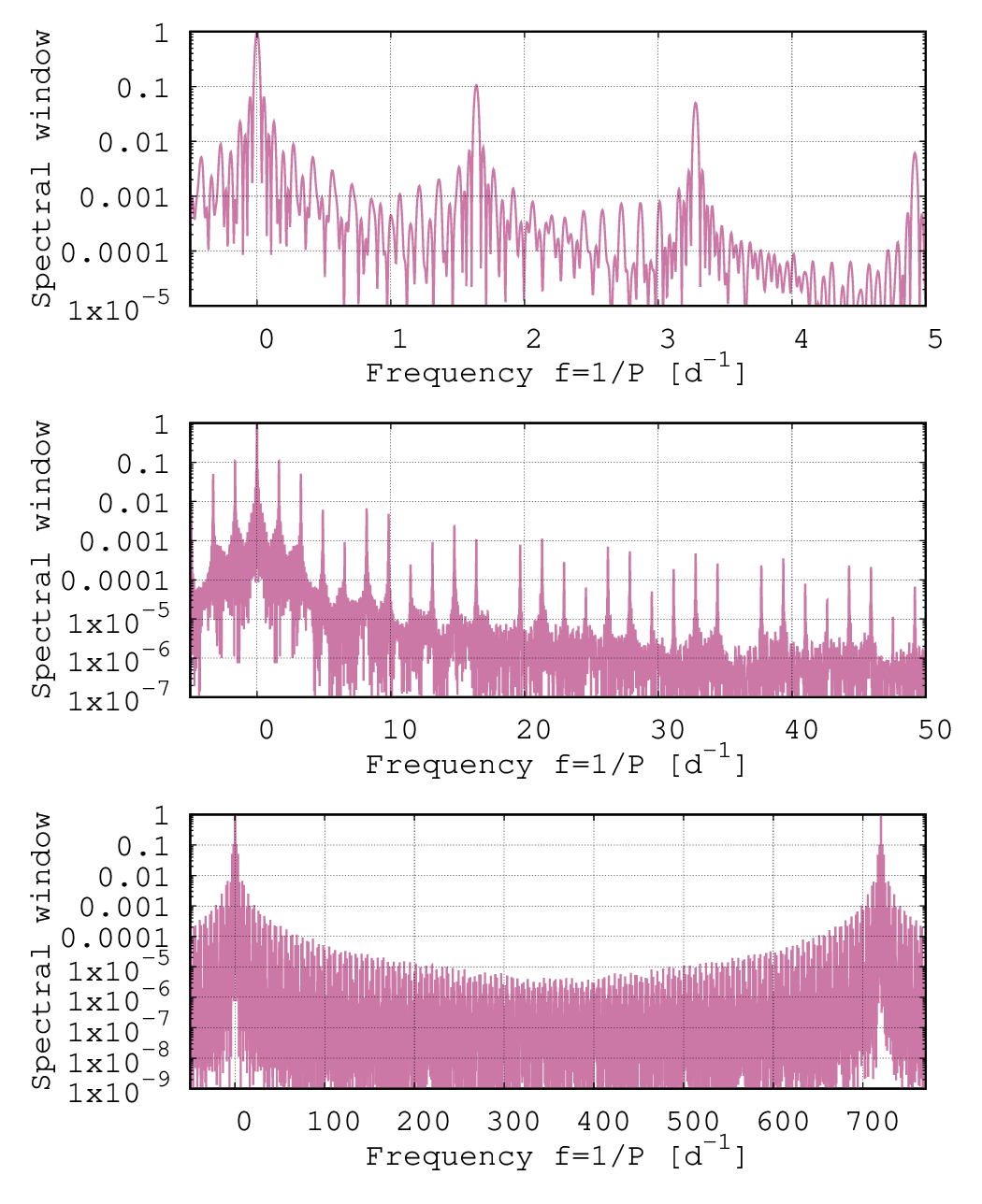}\\
\caption{Spectral window of the
TESS$_{18}$ (left) and TESS$_{58}$ (right) data, for three frequency ranges (from top to
bottom). The (quasi-) Nyquist alias peak at $2f_{\rm Ny} = 720$~d$^{-1}$ is seen in the
bottom panels.}
\label{fig:swnd}
\end{figure*}

TESS data look close to being even with a constant timestep of $\Delta t=2$~min. However,
this is not strictly fulfilled even within just a single sector, because in the middle of
each sector there was a gap, after which the $2$~min grid pattern got shifted. Also, the
OTD set contains holes in place of transits and secondary eclipses. But nevertheless the
spectral window functions of TESS$_{18,58}$ data suggest that these data can be treated as
nearly even, in particular there is a classic aliasing pattern suggesting the existence of
a ``quasi-Nyquist`` frequency $f_{\rm Ny}=1/(2\Delta t)=360$~d$^{-1}$ (see
Fig.~\ref{fig:swnd}).

\subsection{Choosing the TESS data product to process}
\label{sec_select}
\begin{figure*}
\includegraphics[width=\linewidth]{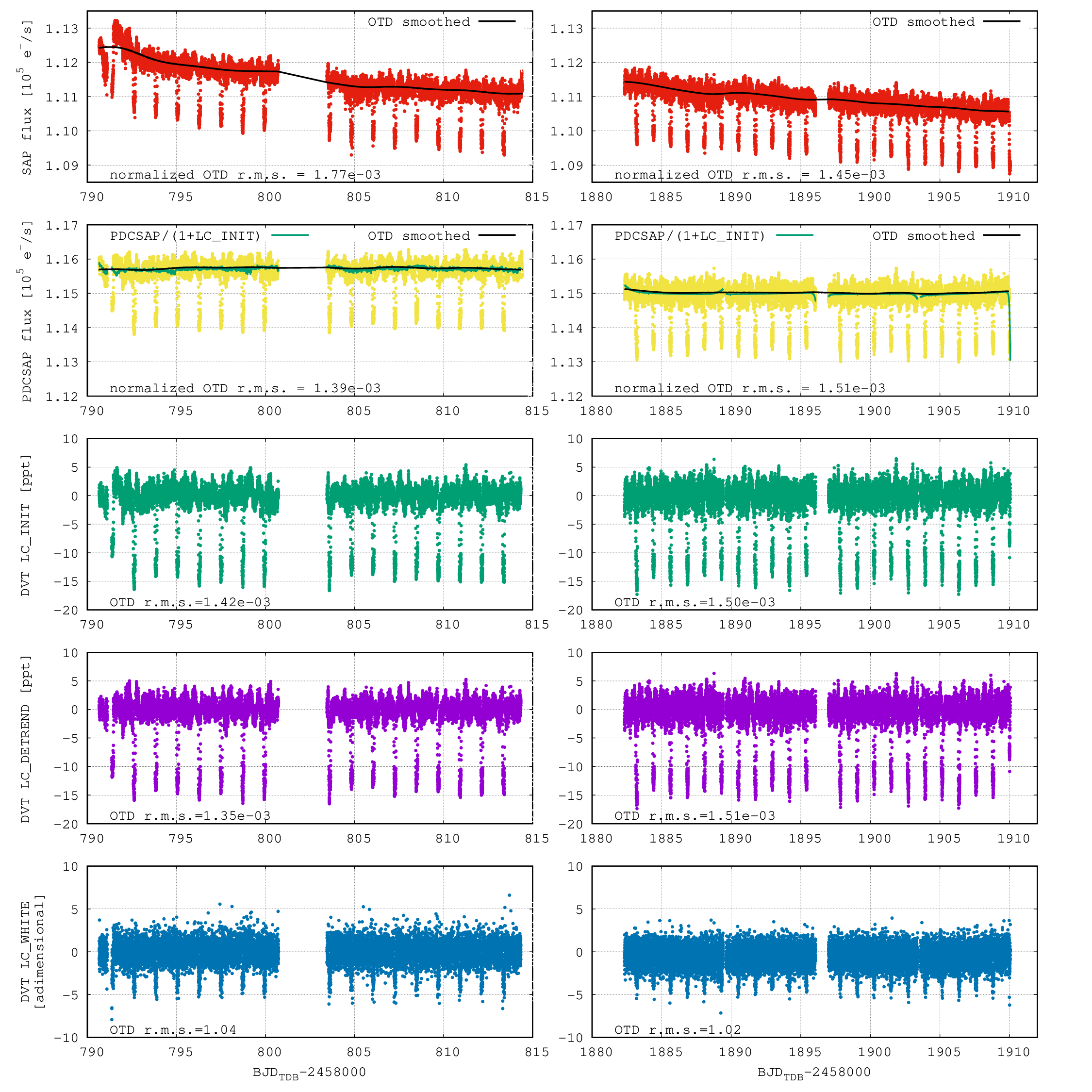}
\caption{Different types of TESS photometric data products of WASP-33, for the Sector~18
(left) and Sector~58 (right); the data types are given in the ordinate labels. Top four
panels additionally show a trend curve (black) obtained by smoothing the OTD with a
Gaussian kernel (sigma of $1$~d). In these four plots we also print the r.m.s. of the OTD
flux divided by the smooth model. All DVT graphs  show the r.m.s. of the corresponding OTD
(which were already normalized) too. The r.m.s. are comparable between each other from top
to the fourth row, while in the bottom plots they are close to unity by whitening.}
\label{fig:data}
\end{figure*}

A yet another important question is the choice of the photometric product for the analysis.
For each target, MAST offers a rich line of photometric data that correspond to different
levels of reduction and processing \citep{Twicken18,Li19,Twicken20,Jenkins20}. In
Fig.~\ref{fig:data} we plot all types of photometric data available for WASP-33.
\citet{vonEssen20} considered two lowest-level types of data, namely SAP (Simple Aperture
Photometry) and PDCSAP (Presearch Data Conditioning SAP), and selected the raw SAP one
motivating this choice by the increased level of correlated noise revealed in PDCSAP. They
estimated this increase by a factor of $\sim 2.4$ using the $\beta$ method by
\citet{Winn08} (their citation to \citealt{CarterWinn09} appears incorrect). It is
simultaneously recognised that WASP-33 stellar pulsations cause larger photometric effect
than possible PDC-generated correlated noise, and in such a case so large increase factor
as $2.4$ looks relatively puzzling. However, the $\beta$ method they used deals with
\emph{relative} noise contributions. The relative portion of a non-white noise (NWN
hereafter) may indeed increase whenever PDC added some, but alternatively this same picture
would appear if PDC \emph{removed} a portion of the white noise (WN hereafter), keeping the
NWN part intact.

Indeed, let us look at the raw SAP data (Fig.~\ref{fig:data}, top panels). They reveal
obvious large-scale trend which is removed by PDC (second pair of panels). However, the PDC
effect is not limited to only detrending. To verify this, we approximated the trend by
smoothing the SAP OTD with a Gaussian kernel (width of $1$~d), then divided it by this
simple model, and computed the r.m.s. of the resulting normalized flux. For the sector~18
data this resulted in $1.73$ parts per thousand (ppt). Applying the same procedure to the
PDCSAP OTD we obtained the r.m.s. of $1.38$~ppt. One can see that additionally to the
trend, the PDC removed $\sim 1.04$~ppt in terms of standard deviation (only in sector 18
data; in sector 58 the r.m.s. appeared nearly identical). Therefore, based on this
preliminary analysis we believe that discarding PDCSAP in favour of the raw SAP data does
not look justified enough, in this particular case at least. Even if PDC did add some extra
noise to the data, it simultaneously removed more.

Beyond the PDC, the DVT (Data Validation Timeseries) are publicly released. They involve
three levels of reduction, which can be refered to by the corresponding column titles in
the DVT files: LC\_INIT, LC\_WHITE, and LC\_DETREND. Briefly, the DVT involve (i) optional
harmonics removal, (ii) edge detrending, (iii) level adjustment, (iv) normalization, (v)
gap unfilling, and (vi) stitching between sectors \citep{Twicken20}. The LC\_INIT data for
WASP-33 (see third pair of plots in Fig.~\ref{fig:data}) have only a minimal difference
from the PDCSAP. This can be verified by considering the graphs of the ratio ${\rm
PDCSAP}/(1+{\rm LC\_INIT})$, which are shown in the second pair of plots in
Fig.~\ref{fig:data}. We can see that LC\_INIT corrects for minor edge-jump effects that are
likely tied to the satellite momentum dumps every few days, which were also synchronized
with pauses made in observations for data download.\footnote{See Data Release Notes at
\url{https://archive.stsci.edu/missions/tess/doc/tess_drn/tess_sector_18_drn25_v02.pdf} and
\url{https://archive.stsci.edu/missions/tess/doc/tess_drn/tess_sector_58_drn83_v02.pdf}} We
believe that this effect is important for our study and should be necessarily removed. To
be honest, its reduction provided with the LC\_INIT data is not always perfect, see e.g.
the end of sector~58 (Fig.~\ref{fig:data}, right, second panel) that was fit badly because
of a partial transit. However, building a more efficient model for this effect would be out
of the scope of current work. That problematic partial transit was removed from our OTD
anyway, and it was not included in the ITD data as well, so the corrupted portion of data
did not affect our analysis.

Concerning the columns LC\_WHITE and LC\_DETREND, they involve more substantial changes to
the flux. It seems these reductions perform at least qualitatively similar to our analysis,
so they will interfere with it, so we selected the DVT LC\_INIT data.

The final notice is that the ``third light'' correction is already applied at the PDC step,
so we do not need to apply it here. There are three known sources that jointly provide
about $2$ per cent extra flux inside the photometric aperture of WASP-33
\citep{vonEssen20}. Their estimation agreed with the TESS-provided one within $\sim 10$ per
cent (the CROWDSAP field stored in the FITS file).

\section{Building the statistically exhaustive model of WASP-33 variability}
\label{sec_analysis}
\subsection{Models}
\label{sec_models}
Here our goal is to construct a statistically ``exhaustive'' WASP-33 model that represents
its OTD so that the residuals do not contain statistically significant periodic (or
quasiperiodic) variations. In this attempt we aimed to apply a periodogram-based approach
somewhat reminiscent of the CLEANest algorithm by \citet{Foster95}. Namely, we represent
the photometric variation as a sum of sinusoidal harmonics:
\begin{equation}
\mu_n(t) = A_0 + \sum_{k=1}^n A_k \cos\left(2\pi f_k (t-T_0) + \varphi_k\right),
\label{mu}
\end{equation}
where $A_k$, $\varphi_k$, and $f_k=1/P_k$ are free parameters.

However, if the model~(\ref{mu}) was used in the white noise treatment then we would likely
need hundreds of harmonics to represent the OTD exhaustively. After extraction of some
initial harmonics ($5$ to $7$ of the most significant ones), the residual periodogram did
not reveal a single clearly dominating peak, looking more similar to a spectrum of banded
NWN.

Alone such an observation is not yet decisive to claim that the data physically involve
NWN. Two alternative models ``a lot of harmonics + WN'' and ``fewer harmonics + NWN'' may
appear formally indistinguishable, i.e. statistically equivalent in how well they describe
the data. But even in this case the NWN model may offer a great reduction in the number of
model components (and parameters), so we can profit from a mathematically more simple
$\mu$.

To take the NWN into account, we adopted an approach of Gaussian process (GP) fitting,
introduced in \citep{Baluev11}. In this approach we should define a model for the NWN
correlation function, which we now set to:
\begin{align}
k_{n'}(\Delta t_{ij}) &= \left(\sigma_{{\rm TESS},i}^2+\sigma_0^2\right) \delta_{ij} + \sum_{k=1}^{n'} \sigma_k^2 {\rm e}^{-\frac{|\Delta t_{ij}|}{\tau_k}} \cos 2\pi f'_k \Delta t_{ij}, \nonumber\\
&\phantom{=} \Delta t_{ij} = t_i - t_j,
\label{corr}
\end{align}
where $\sigma_{{\rm TESS},i}$ is TESS-provided photometric uncertainty at $t_i$,
$\delta_{ij}$ is Kronecker delta, while $\sigma_k$, $\tau_k$, and $f'_k=1/P'_k$ are free
parameters. This model involves the WN portion ($\sigma_0$) and $n'$ independent
quasiperiodic components (purely NWN part). By a visual look at the periodograms (e.g. see
Fig.~\ref{fig:prdgs} discussed below) we noticed three possible spectrum bands that are
concentrated near the frequencies $f'_1\simeq 0$, $f'_2\simeq 10$~d$^{-1}$, and $f'_3\simeq
20$~d$^{-1}$. We set these frequencies as starting values for the GP fitting, thus assuming
$n'=3$.

The models~(\ref{mu}) and~(\ref{corr}) can be fitted jointly using the maximum likelihood
approach, assuming that noise is a Gaussian random process \citep{Baluev11,Baluev13a}. It
is based on minimizing the function
\begin{eqnarray}
f(\btheta, \boldeta) = \log\det\mathbfss V(\boldeta) + \frac{1}{\gamma}\,
{\bmath r}^{\mathrm T}(\btheta)\, \mathbfss V^{-1}(\boldeta)\, \bmath r(\btheta) \; \stackrel{\btheta,\boldeta}{\longmapsto} \; \min, \nonumber\\
\bmath r = \bmath x - \bmu(\bmath t,\btheta), \quad \gamma = 1-\frac{\dim\btheta}{\dim\bmath x},
\label{objf}
\end{eqnarray}
where we have (i) input data including observation times $\bmath t$ and the flux $\bmath
x$, (ii) the models $\bmu$ being one of~(\ref{mu}) and $\mathbfss V$ which elements are
filled in with~(\ref{corr}), and (iii) vectors of fittable parameters $\btheta$ (which
includes $A_k$, $\varphi_k$, and $f_k$) and $\boldeta$ (which includes $\sigma_k^2$,
$\tau_k$, and $f_k'$). The correction factor $\gamma$ is necessary to reduce the bias in
$\boldeta$.

\subsection{Algorithm}
We run a sequential periodogram algorithm according to the following scheme. We started
from $k_3$ as specified above, and from $\mu_2$ involving two harmonics with fixed
frequencies as per \citet{vonEssen20}: $f_1=P_b$ (orbital phase variation) and $f_2=P_b/2$
(ellipsoidal variation). After that the model~(\ref{mu}) can be built up iteratively by
adding harmonics that refer to the tallest peak in the subsequent residual periodogram.
These periodograms must take the NWN into account via~(\ref{corr}). For that we used the
formalism of the likelihood-ratio periodogram (LRP) introduced by
\citet{Baluev08b,Baluev13a} and implemented in the {\sc PlanetPack} software
\citep{Baluev13c,Baluev18c}. The LRP considers two rival models of the data, a base one
($\mu_n$ on $n$th iteration) and an alternative one involving an extra sinusoid (i.e.
$\mu_{n+1}$). The likelihood ratio statistic based on~(\ref{objf}) is then computed for
these models and its logarithm is plotted against the frequency of the last added sinusoid.
Such a graph represents our LRP that generalises the classic \citet{Lomb76}~--
\citet{Scargle82} periodogram. The LRP can be computed for the classic WN case as well as
for the NWN model~(\ref{corr}), by defining the likelihood function (namely, the model
$\mathbfss V(\boldeta)$) appropriately. After each iteration all free parameters of so far
detected harmonics are refined by refitting with the new model $\mu$.

Very importantly, the spectral window in Fig.~\ref{fig:swnd} does not demonstrate large
side peaks (in the Nyquist range at least). The height of the tallest side peak is $\simeq
0.1$. Therefore, the aliasing appears insignificant in our task, and we may safely assume
that the maximum periodogram peak corresponds to the true (rather than alias) frequency.
This makes all the analysis easier, for example it is unnecessary to care about the order
in which the harmonics are extracted (which could be important otherwise).

However, dealing with NWN periodograms implies numerous NWN model refitting, which is
computationally demanding. The likelihood function involves operations on very large
covariance matrices ($\sim 10^4 \times 10^4$), rendering the LRP computation too slow for
being accomplished in full. But we tried to bypass this limitation using the following
approach. On each iteration we computed, at first, the fast WN-only version of the LRP, and
determined its $256$ tallest peaks. Then for each of these peaks we run an NWN
maximum-likelihood fitting, starting it from the associated WN best fit. This allowed us to
compute heights of these same peaks in the NWN treatment, and then to select the tallest
one among them.

Such an approach is potentially vulnerable because it may, in theory, miss the actual
best-significance harmonic. Even $256$ peaks that appeared tallest with the WN model do not
necessarily contain the one that should appear tallest in the NWN case. However, the
probability of such a condition decreases when the number of peaks being inspected grows,
so the validity of the results can be tested by a probe increase of that number (in the end
of iterations sequence).

We launched our harmonics detection iterations assuming the frequency range from $f=0$ to
$f=50$~d$^{-1}$, which is \emph{a priori} reasonable for $\delta$~Scuti stars, and also
contained all apparently noticeable periodogram power. As we specified above, the NWN
model~(\ref{corr}) was initially set to have $n'=3$ terms, but somewhere in the middle of
the sequence the third NWN term (with $f'\simeq 20$~d$^{-1}$) appeared, basically,
decomposed into harmonics. We started to obtain various degeneracy issues regarding this
term, for example it was trying to either model just a single sinusoid ($\tau\to 0$), or to
mimic a portion of the white noise ($\tau\to\infty$), or to model a portion of another NWN
term. When this behaviour started, we manually removed this third term from the NWN model,
and continued the sequence further with $k_2(\Delta t)$ instead of $k_3(\Delta t)$.
Finally, when no significant harmonics left (in the NWN treatment), we reprocessed the last
iteration inspecting $1024$ tallest peaks instead of just $256$, and in the entire Nyquist
range (max frequency of $360$~d$^{-1}$ instead of $50$~d$^{-1}$). This did not result in
detection of additional harmonics, ensuring that we did not miss any.

\begin{figure*}
\def\myscalehere{\linewidth*\real{0.90}}
\includegraphics[trim={0.45cm 1.30cm 0.55cm 0.35cm},clip,height=0.1806\myscalehere]{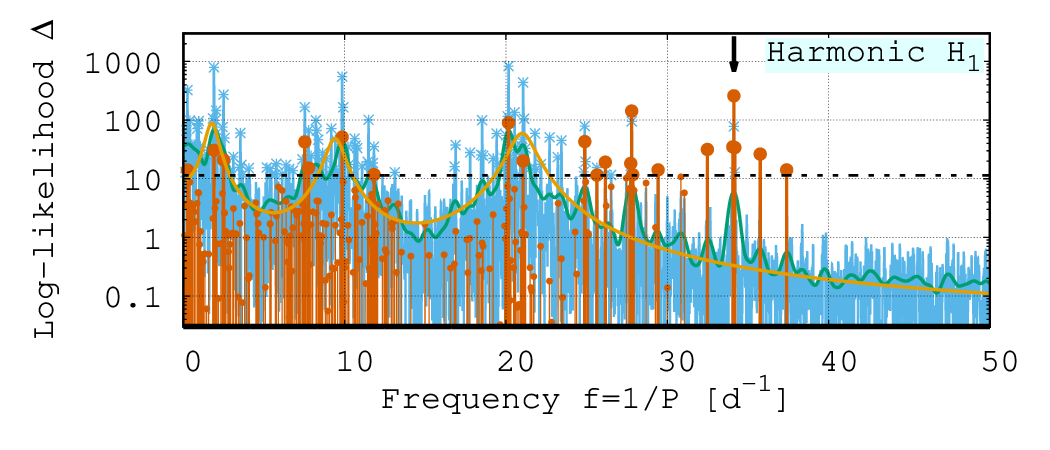}
\includegraphics[trim={1.30cm 1.30cm 0.55cm 0.35cm},clip,height=0.1806\myscalehere]{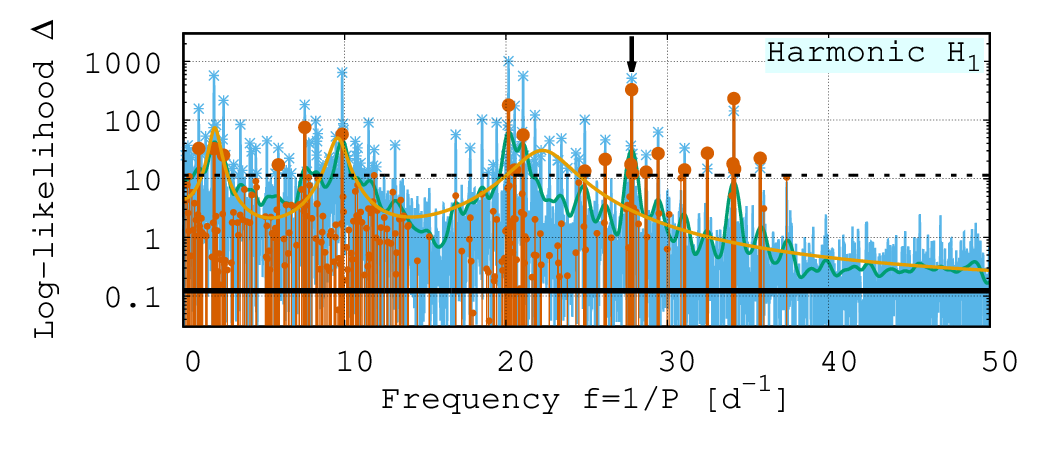}\\
\includegraphics[trim={0.45cm 1.30cm 0.55cm 0.35cm},clip,height=0.1806\myscalehere]{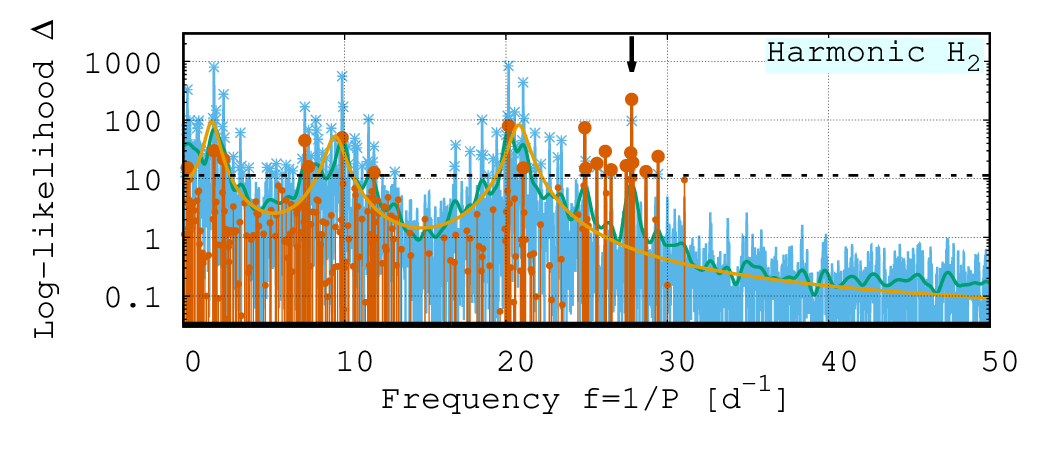}
\includegraphics[trim={1.30cm 1.30cm 0.55cm 0.35cm},clip,height=0.1806\myscalehere]{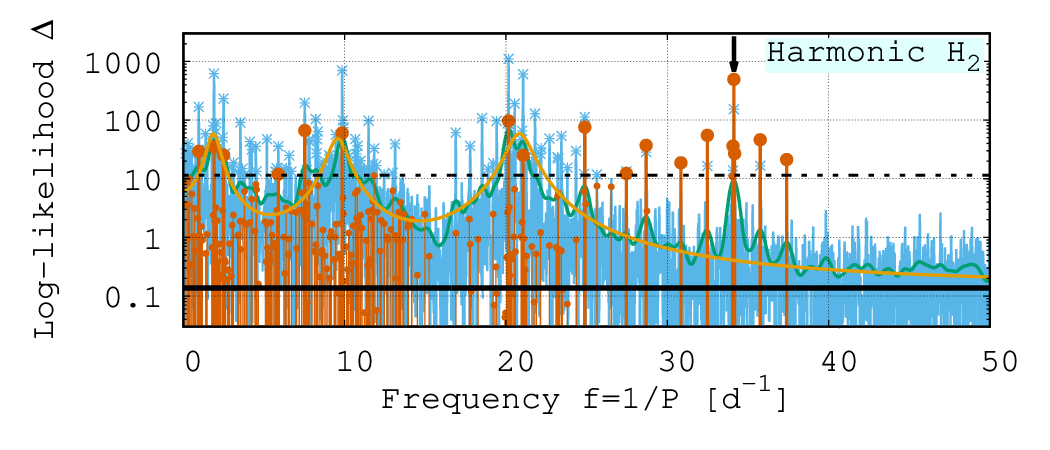}\\
\includegraphics[trim={0.45cm 1.30cm 0.55cm 0.35cm},clip,height=0.1806\myscalehere]{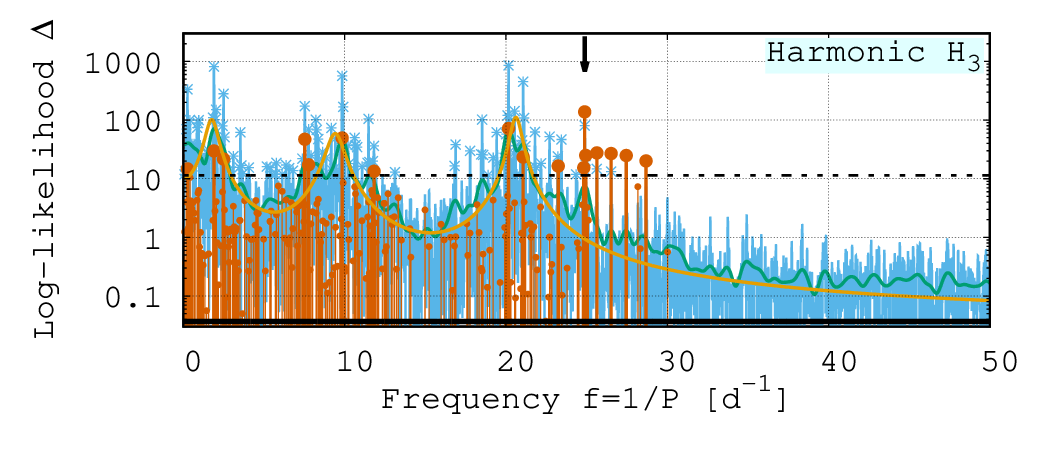}
\includegraphics[trim={1.30cm 1.30cm 0.55cm 0.35cm},clip,height=0.1806\myscalehere]{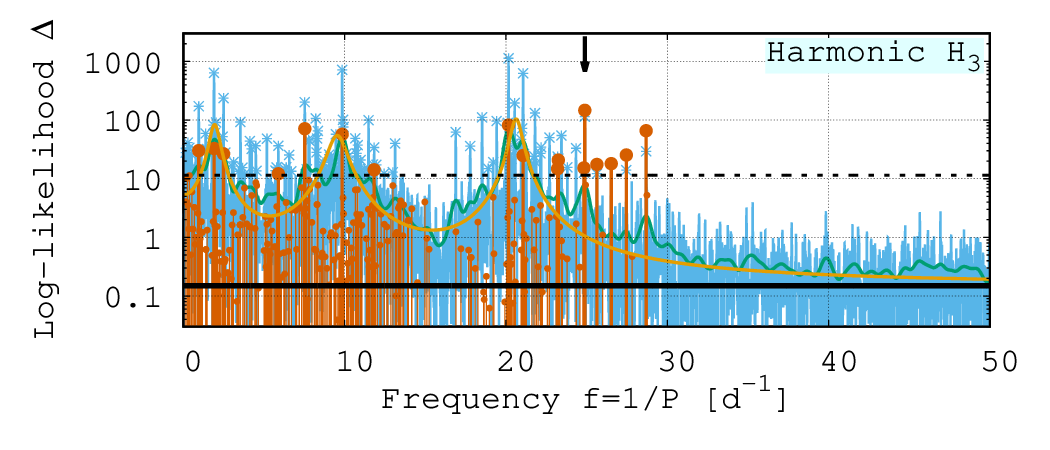}\\
\includegraphics[trim={0.45cm 0.00cm 0.55cm 0.35cm},clip,height=0.22\myscalehere]{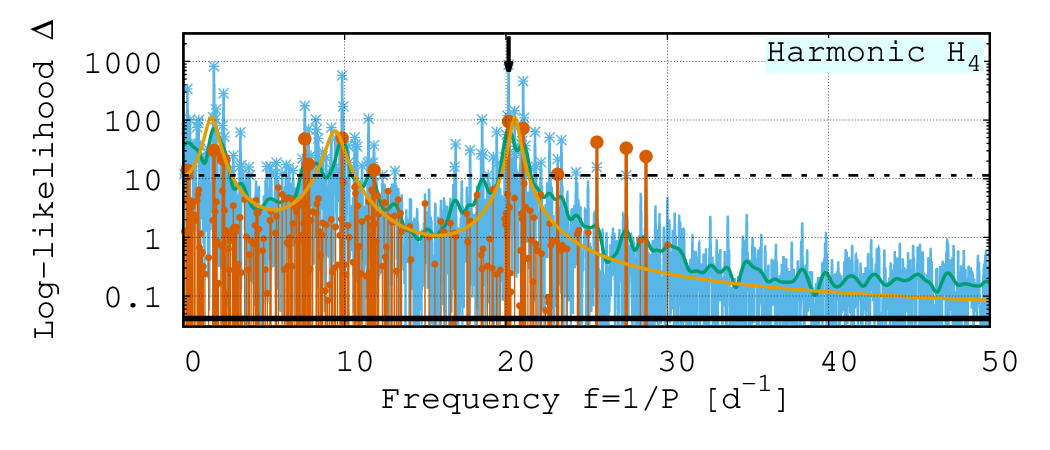}
\includegraphics[trim={1.30cm 0.00cm 0.55cm 0.35cm},clip,height=0.22\myscalehere]{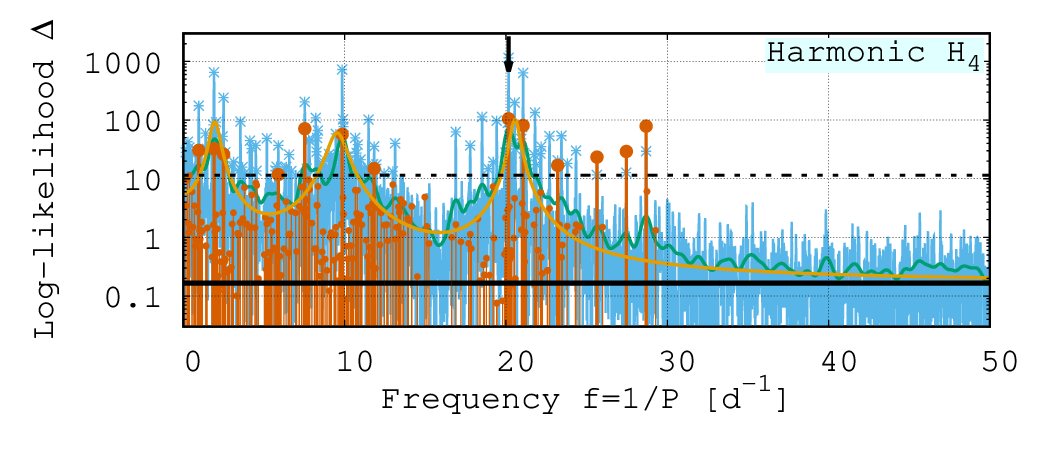}\\
\includegraphics[trim={0 0.5cm 0 0.45cm},clip,width=0.49\myscalehere]{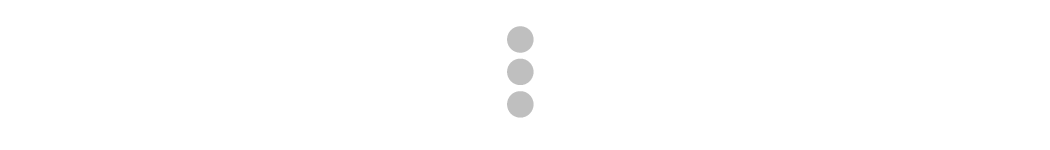}
\includegraphics[trim={0 0.5cm 0 0.45cm},clip,width=0.49\myscalehere]{prdgDOTS.eps}\\
\includegraphics[trim={0.45cm 0.00cm 0.55cm 0.35cm},clip,height=0.22\myscalehere]{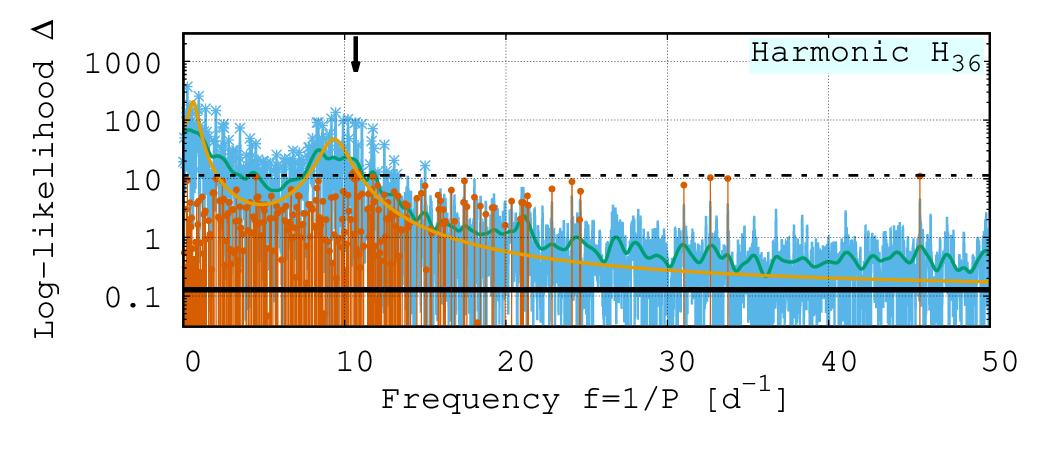}
\includegraphics[trim={1.30cm 0.00cm 0.55cm 0.35cm},clip,height=0.22\myscalehere]{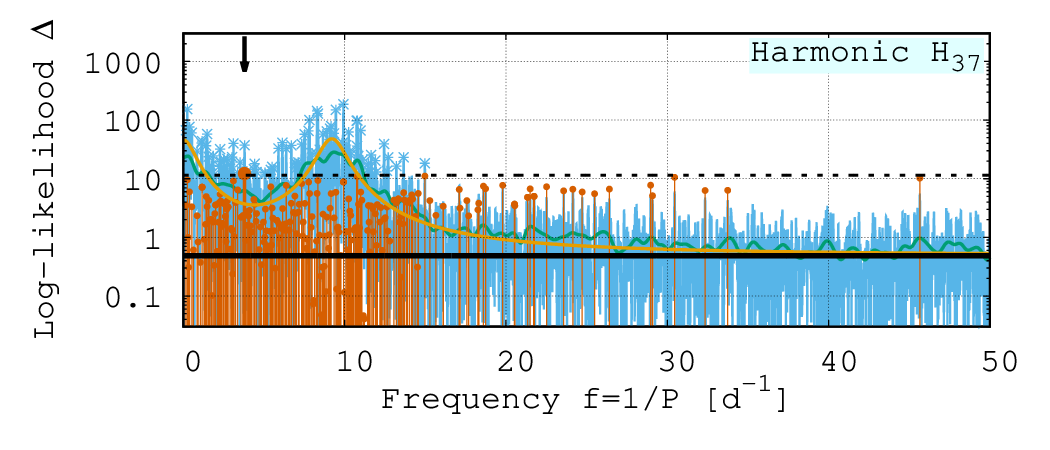}\\
\includegraphics[trim={0.45cm 0.00cm 0.55cm 0.35cm},clip,height=0.22\myscalehere]{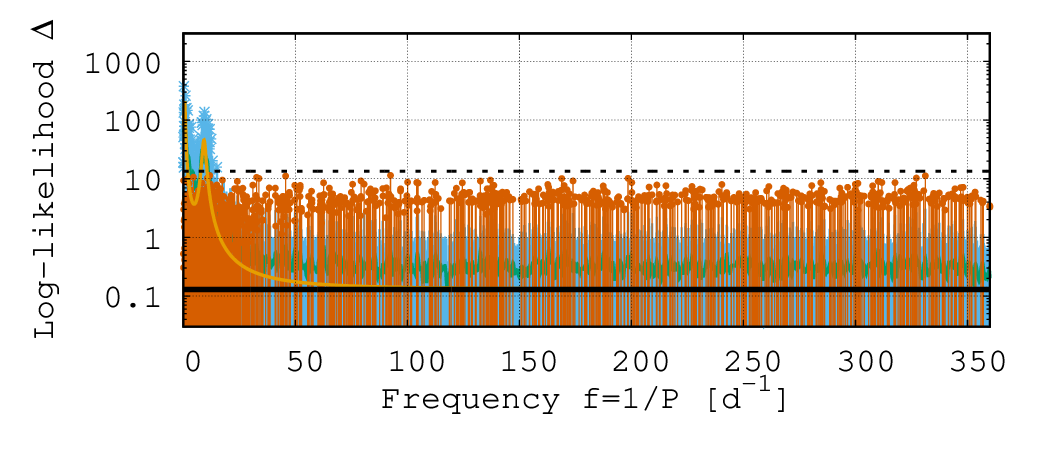}
\includegraphics[trim={1.30cm 0.00cm 0.55cm 0.35cm},clip,height=0.22\myscalehere]{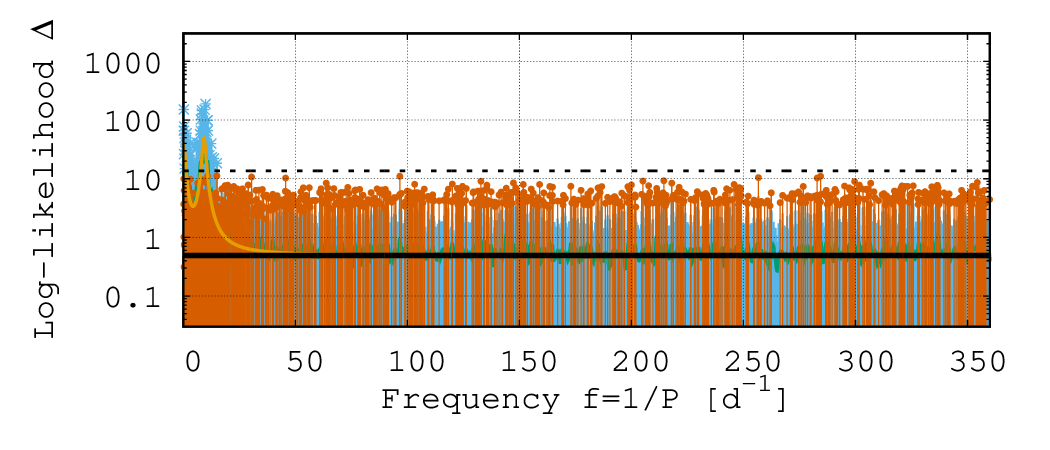}\\
\includegraphics[trim={1.00cm 0.80cm 1.25cm 0.70cm},clip,width=\myscalehere]{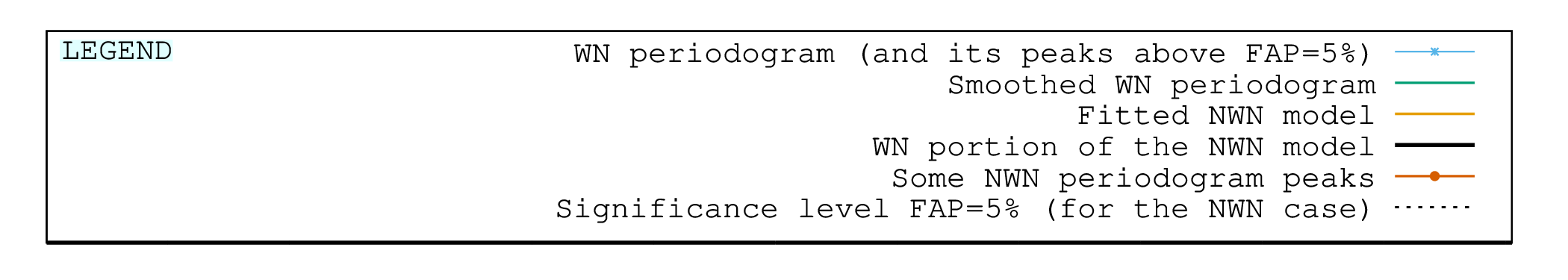}\\
\caption{Periodogram analysis and extraction of the harmonics. The full set of figures is
available in the online-only supplement.}
\label{fig:prdgs}
\end{figure*}

This algorithm was applied to TESS$_{18}$ and TESS$_{58}$ data separately, which resulted
in quite similar iterations sequences and in similar results. Some graphs illustrating the
initial and final iterations are shown in Fig.~\ref{fig:prdgs}. Each panel demonstrates the
WN periodograms (raw as well as smoothed one), the theoretic power spectrum of the current
best fitting NWN model based on~(\ref{corr}), and the set of NWN peaks that were dealt with
on this iteration. On the last iteration, when we inspected $1024$ peaks, those peaks are
covering the entire Nyquist range without significant holes or blind segments, so it is
very unlikely that some statistically significant harmonics escaped our analysis. Please
also notice remarkable similarity between our fitted NWN power spectrum and the smoothed WN
periodogram.

This algorithm belongs to a frequentist type, and is based, in its core, on the likelihood
ratio test. We did not try methods of a Bayesian type here, because it was shown by
\citet{Baluev22} that in the task of sinusoidal signal detection, without significant
aliasing at least, detection efficiency of a Bayesian analysis (measured by certain formal
metric) appears practically identical to the usual non-Bayesian periodograms.

\subsection{Detection threshold}
An explanation should be given regarding how we computed the statistical significance of
periodogram peaks in the NWN treatment. As explained by \citet{Baluev08b}, for the WN type
of the LRP we may use the following approximation of the False Alarm Probability (FAP) from
\citep{Baluev08a}:
\begin{equation}
{\rm FAP}(z) \stackrel{z\gg 1}{\simeq} W {\rm e}^{-z} \sqrt z, \quad W = (f_{\max}-f_{\min}) T_{\rm eff},
\label{fap}
\end{equation}
where $z$ is the height of the tallest peak, and $T_{\rm eff}$ is an effective time span
proportional to the sample variance of the timings $t_i$. As argued (without proofs) in
\citep{Baluev14a}, the approximate formula~(\ref{fap}) remains applicable for NWN models as
well, but with $T_{\rm eff}$ somewhat differently determined. \citet{Delisle20} developed
this study further, partly confirming this conclusion and providing a more general
definition for $T_{\rm eff}$. However, in our task it appeared that $T_{\rm eff}$ always
remains practically identical, regardless of the noise model or which approximation to
adopt. We obtained $T_{\rm eff}$ about $24$~d for TESS$_{18}$ and about $28$~d for
TESS$_{58}$.

In this study we aimed to obtain a more accurate final model of the WASP-33 variability,
possibly even by the cost of larger probability of false positive harmonics. So we adopted
a relatively mild FAP threshold of $5\%$ (the iterations continued until the right hand
side of~(\ref{fap}) exceeded $0.05$).

\subsection{Results}
Table~\ref{tab:harmonics} lists all the harmonics that we detected in TESS$_{18}$ and in
TESS$_{58}$.\footnote{Except for the two harmonics with fixed periods $P_b$ and $P_b/2$.
Those are not shown in Table~\ref{tab:harmonics} for clarity, and are discussed
separately.} It is highly important that both these sectors revealed very similar sets of
the harmonics. Nearly all harmonics can be cross-identified between TESS$_{18}$ and
TESS$_{58}$ by comparing their frequencies (honouring uncertainties). Usually, the
amplitudes of the cross-identified harmonics appeared similar as well, although some of
$A_k$ demonstrated big changes between TESS$_{18}$ and TESS$_{58}$. It is impossible to
conclusively compare the phases of these harmonics, because of too large uncertainty
accumulated over three years passed between these TESS sectors. In general, one may
conclude that WASP-33 variability pattern demonstrates remarkable, though not entirely
perfect, stability over the timescale of a few years at least.

\begin{table*}
\caption{Estimated harmonics parameters.}
\label{tab:harmonics}
%\footnotesize
\begin{center}
\begin{tabular}{llllllll}
\hline
         &          &\multicolumn{3}{c}{
                      TESS$_{18}$ ($T_0=2458800$)}   &
                     \multicolumn{3}{c}{
                      TESS$_{58}$ ($T_0=2459895$)}   \\
\multicolumn{2}{c}{IDs}&$P$~[d]        &$A$~[ppm]     &$\varphi$~[$^\circ$]&$P$~[d]      &$A$~[ppm]     &$\varphi$~[$^\circ$]\\
\hline
$H_1$    &          &                  &              &               & $0.0213065 (17)$ & $ 41.7(8.0)$ & $328 (11)  $ \\
$H_2$    & $F_{10}$ & $0.02930411(56)$ & $228.3(6.8)$ &  $297.9(1.8)$ & $0.02930358(45)$ & $312.5(8.3)$ & $279.2(1.5)$ \\
$H_3$    & $F_{28}$ & $0.0328267 (44)$ & $ 38.0(7.0)$ &  $306 (11)  $ &                  &              &              \\
$H_4$    & $F_{26}$ & $0.0331895 (39)$ & $ 44.5(7.1)$ &  $ 37.1(9.8)$ & $0.0331986 (43)$ & $ 42.4(8.5)$ & $155 (12)  $ \\
$H_5$    & $F_{22}$ & $0.0348597 (28)$ & $ 69.6(7.2)$ &  $206.2(6.4)$ & $0.0348553 (16)$ & $130.6(8.7)$ & $186.8(3.8)$ \\
$H_6$    & $F_8$    & $0.03597781(78)$ & $267.3(7.3)$ &  $317.9(1.7)$ & $0.03597797(38)$ & $580.6(8.7)$ & $ 93.27 (87)$\\
$H_7$    & $F_{18}$ & $0.0364189 (26)$ & $ 84.0(7.3)$ &  $180.1(5.4)$ & $0.0364150 (26)$ & $ 86.0(8.7)$ & $315.9(5.9)$ \\
$H_8$    &          &                  &              &               & $0.0373459 (57)$ & $ 42.3(8.8)$ & $154 (12)  $ \\
$H_9$    & $F_{14}$ & $0.0390044 (23)$ & $111.5(7.5)$ &  $198.8(4.2)$ & $0.0390015 (30)$ & $ 87.9(8.9)$ & $214.6(5.9)$ \\
$H_{10}$ & $F_9$    & $0.0401869 (11)$ & $253.0(7.8)$ &  $ 11.5(1.9)$ & $0.0401856 (11)$ & $258.3(9.1)$ & $118.1(2.1)$ \\
$H_{11}$ &          & $0.0406911 (67)$ & $ 43.5(8.0)$ &  $  9 (11)  $ &                  &              &              \\
$H_{12}$ & $F_{16}$ & $0.0430933 (35)$ & $102.1(8.2)$ &  $258.8(5.1)$ & $0.0430910 (30)$ & $114.8(9.4)$ & $ 55.1(4.8)$ \\
$H_{13}$ & $F_{27}$ & $0.0440461 (85)$ & $ 41.3(8.7)$ &  $121 (13)  $ & $0.0440554 (71)$ & $ 51 (10)  $ & $221 (11)  $ \\
$H_{14}$ & $F_{20}$ & $0.0451501 (44)$ & $ 90.3(8.5)$ &  $224.4(6.0)$ & $0.0451491 (44)$ & $ 91.3(9.9)$ & $245.1(6.1)$ \\
$H_{15}$ &          & $0.0454574 (96)$ & $ 40.6(8.6)$ &  $103 (13)  $ &                  &              &              \\
$H_{16}$ &          & $0.0457087 (92)$ & $ 43.5(8.7)$ &  $ 60 (12)  $ & $0.0457113 (72)$ & $ 56.2(9.8)$ & $194 (10)$   \\
$H_{17}$ & $F_{19}$ & $0.0460159 (54)$ & $ 76.8(8.8)$ &  $261.9(7.0)$ & $0.0460004 (57)$ & $ 72.1(9.8)$ & $ 73.5(7.8)$ \\
$H_{18}$ &          & $0.0465026 (83)$ & $ 53.0(8.8)$ &  $296 (10)  $ &                  &              &              \\
$H_{19}$ & $F_{24}$ & $0.0469927 (71)$ & $ 61.5(8.8)$ &  $164.0(8.7)$ & $0.0469969 (49)$ & $ 85.6(9.7)$ & $268.1(6.5)$ \\
$H_{20}$ & $F_4$    & $0.04747567(95)$ & $617 (11)  $ &  $335.46 (96)$& $0.04747626(78)$ & $618 (11)  $ & $161.2(1.0)$ \\
$H_{21}$ & $F_{11}$ & $0.0476547 (89)$ & $167 (22)  $ &  $ 54.4(4.4)$ & $0.0476614 (36)$ & $190 (11)  $ & $ 42.4(3.4)$ \\
$H_{22}$ &          & $0.047747  (12)$ & $119 (21)  $ &  $ 86.7(5.9)$ & $0.0477543 (62)$ & $102 (11)  $ & $101.6(6.5)$ \\
$H_{23}$ & $F_6$    & $0.0486902 (15)$ & $335.7(8.9)$ &  $259.1(1.7)$ & $0.0486897 (15)$ & $330 (10)  $ & $125.5(1.8)$ \\
$H_{24}$ & $F_1$    & $0.04959774(65)$ & $780.8(9.1)$ &  $  1.13 (72)$& $0.04959762(62)$ & $786 (10)  $ & $265.35(74)$ \\
$H_{25}$ & $F_{17}$ & $0.0500732 (53)$ & $100.1(9.1)$ &  $186.5(5.6)$ & $0.0500786 (41)$ & $123 (10)  $ & $ 79.8(4.8)$ \\
$H_{26}$ & $F_{15}$ & $0.0520673 (53)$ & $105.9(9.3)$ &  $208.5(5.5)$ & $0.0520599 (50)$ & $110 (10)  $ & $229.6(5.4)$ \\
$H_{27}$ & $F_{23}$ & $0.0548976 (91)$ & $ 73 (10)  $ &  $ 41.1(8.4)$ & $0.054928  (11)$ & $ 60 (11)  $ & $228 (10)  $ \\
$H_{28}$ &          &                  &              &               & $0.077011  (20)$ & $100 (18)  $ & $227 (10)  $ \\
$H_{29}$ & $F_{13}$ & $0.084545  (23)$ & $151 (22)  $ &  $184.2(9.0)$ & $0.084530  (21)$ & $158 (23)  $ & $357.8(8.3)$ \\
$H_{30}$ &          &                  &              &               & $0.085006  (30)$ & $110 (23)  $ & $ 68 (12)  $ \\
$H_{31}$ & $F_{12}$ & $0.092850  (34)$ & $190 (33)  $ &  $ 26 (10)  $ & $0.092665  (33)$ & $166 (32)  $ & $  0 (11)  $ \\
$H_{32}$ &          & $0.093638  (44)$ & $157 (34)  $ &  $ 85 (13)  $ & $0.093750  (29)$ & $201 (34)  $ & $ 78.1(9.7)$ \\
$H_{33}$ & $F_3$    & $0.101556  (19)$ & $583 (48)  $ &  $ 67.6(5.1)$ & $0.101567  (15)$ & $638 (47)  $ & $141.5(4.3)$ \\
$H_{34}$ &          & $0.128854  (66)$ & $191 (34)  $ &  $237 (11)  $ &                  &              &              \\
$H_{35}$ & $F_7$    & $0.132814  (40)$ & $301 (31)  $ &  $226.6(6.3)$ & $0.132831  (30)$ & $346 (30)  $ & $338.9(5.1)$ \\
$H_{36}$ &          &                  &              &               & $0.22254   (17)$ & $112 (19)  $ & $126 (10)  $ \\
$H_{37}$ &          & $0.26299   (26)$ & $139 (23)  $ &  $216 (10)  $ & $0.26367   (29)$ & $ 92 (20)  $ & $  8 (12)  $ \\
$H_{38}$ &          & $0.32205   (42)$ & $138 (25)  $ &  $114 (11)  $ &                  &              &              \\
$H_{39}$ & $F_5$    & $0.40112   (28)$ & $384 (28)  $ &  $358.4(4.8)$ & $0.40099   (17)$ & $427 (22)  $ & $349.3(3.1)$ \\
$H_{40}$ &          & $0.47653   (65)$ & $259 (33)  $ &  $ 81.9(7.8)$ & $0.47670   (67)$ & $171 (26)  $ & $127.1(8.3)$ \\
$H_{41}$ &          &                  &              &               & $0.49322   (64)$ & $194 (26)  $ & $140.6(7.5)$ \\
$H_{42}$ & $F_2$    & $0.52602   (31)$ & $713 (36)  $ &  $233.2(3.1)$ & $0.52575   (21)$ & $652 (26)  $ & $316.7(2.3)$ \\
$H_{43}$ &          &                  &              &               & $1.0516    (26)$ & $321 (39)  $ & $188.2(7.1)$ \\
\hline
\end{tabular}\\
\begin{minipage}{13.5cm}
\emph{NB (i)\phantom{ }} Uncertainties are given in parentheses after the estimation, in
the unit of the latter's last significant digit. This applies to all similar tables
below.\\
\emph{NB (ii)} Harmonics labelled with $F_k$ (col.~2) were cross-identified with those reported by
\citet{vonEssen20}; this label is the same as used in that paper.
\end{minipage}
\end{center}
\end{table*}

\begin{figure*}
\includegraphics[width=\linewidth]{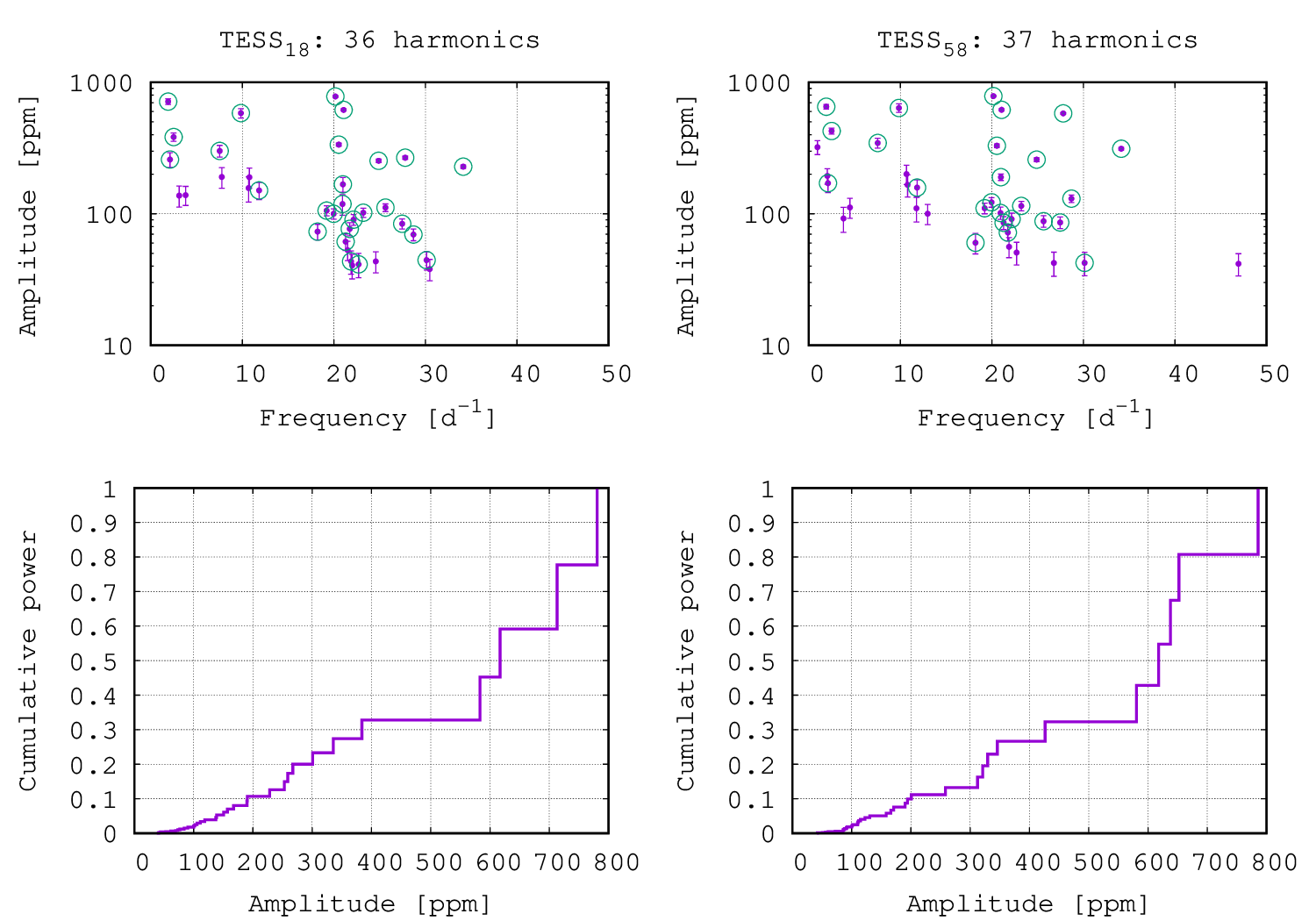}
\caption{All the detected harmonics in the frequency~-- amplitude plane and their
contributions to the total cumulative power.}
\label{fig:harm}
\end{figure*}

In Fig.~\ref{fig:harm}, top panels, we plot all the detected harmonics in the frequency~--
amplitude plane (with uncertainties). Harmonics that were cross-identified by their
frequencies are additionally labelled with circles. We can see that in a given TESS sector
just a few harmonics miss a counterpart in the another one. Such cases might indicate that
the given harmonic is a false positive or vice versa, that the counterpart harmonic failed
to overcome the $\rm FAP=5\%$ threshold level.

In the bottom panels of Fig.~\ref{fig:harm} we plot the cumulative function of the
harmonics power $A_k^2/2$ against the amplitude $A_k$. From these graphs we can see that
$\sim 70\%$ of the total power (variance) of the compound harmonics signal is contained in
just $4$ or $5$ components of largest amplitude ($A_k\gtrsim 600$~ppm), while the remaining
$\sim 30\%$ of the total power is distributed among the rest ($A_k\lesssim 400$~ppm). The
total power of the compound signal corresponds to $\sim 1200$~ppm for TESS$_{18}$ and
TESS$_{58}$ both.

\begin{table}
\caption{Estimated noise parameters.}
\label{tab:NWN}
\small
\begin{center}
\begin{tabular}{llll}
\hline
\multicolumn{2}{c}{Parameter}         & TESS$_{18}$      & TESS$_{58}$     \\
\hline
$\sigma_0$   & [ppm]  & $260.4(5.3)$  & $544.6(4.9)$ \\
$\sigma_1$   & [ppm]  & $525(47)$     & $294(23)$    \\
$\tau_1$     & [d]    & $0.52(13)$    & $0.230(98)$  \\
$P'_1$       & [d]    & $1.65(23)$    & $\infty(\gtrsim 1.7)$\\
$\sigma_2$   & [ppm]  & $426(23)$     & $470(22)$    \\
$\tau_2$     & [d]    & $0.182(26)$   & $0.182(21)$  \\
$P'_2$       & [d]    & $0.1085(15)$  & $0.1084(13)$ \\
\hline
\end{tabular}
\end{center}
\end{table}

Table~\ref{tab:NWN} contains the best fitting NWN parameters of the model~(\ref{corr}). The
total noise variance corresponds to $\sim 700$~ppm and appears nearly the same for TESS$_{18}$
and TESS$_{58}$. Within it, the second NWN term appears nearly the same, although the first
one reveals a significant change between the sectors, as well as the WN term does. But
because of the stable total variance, this difference looks like a power exchange between
the WN and the first NWN term over time. The total model variance becomes therefore $\sim
1400$~ppm for the both TESS sectors, and the variance of these data themselves matches this
level closely, confirming that our models are consistent and close to being statistically
exhaustive.

\begin{figure*}
\includegraphics[trim={0.45cm 0 0.55cm 0.35cm},clip,height=0.53\linewidth]{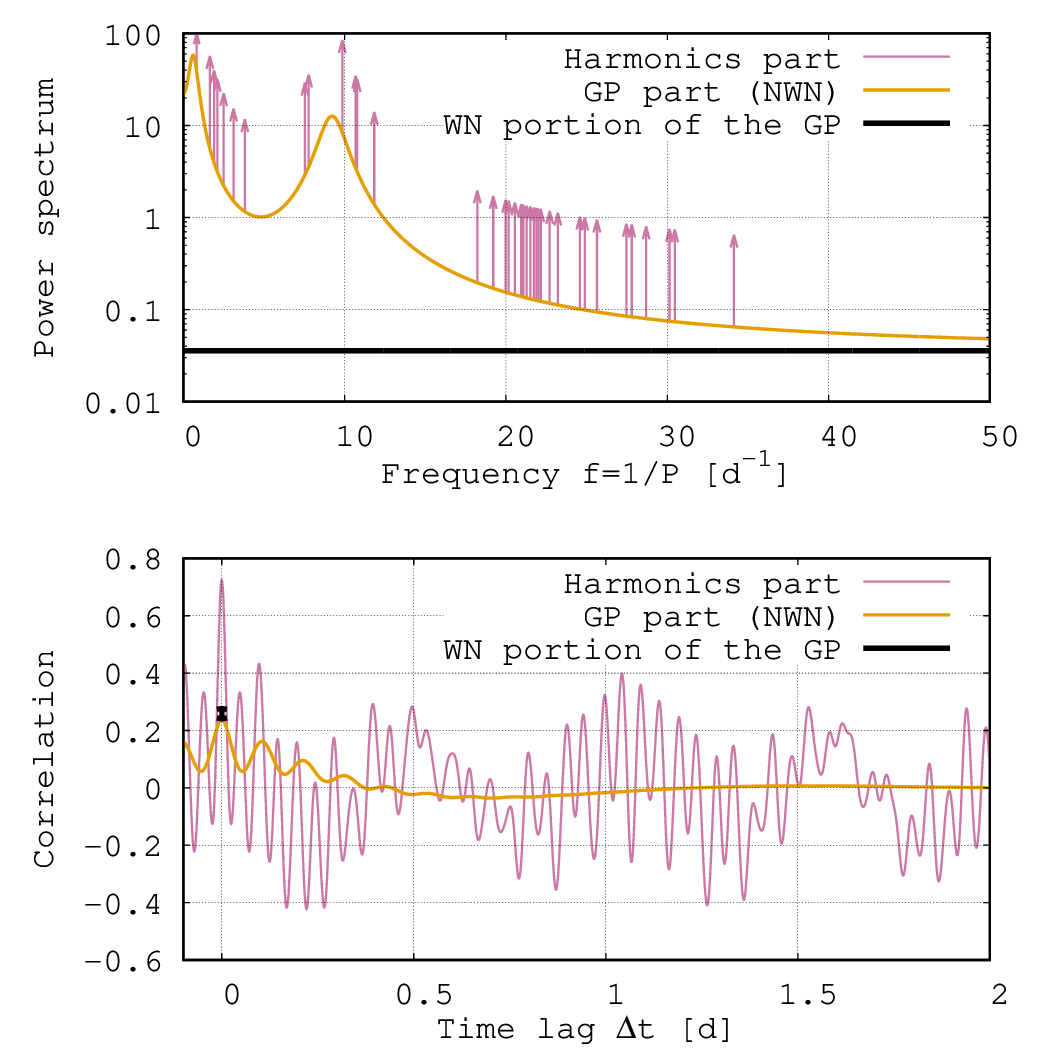}
\includegraphics[trim={1.35cm 0 0.55cm 0.35cm},clip,height=0.53\linewidth]{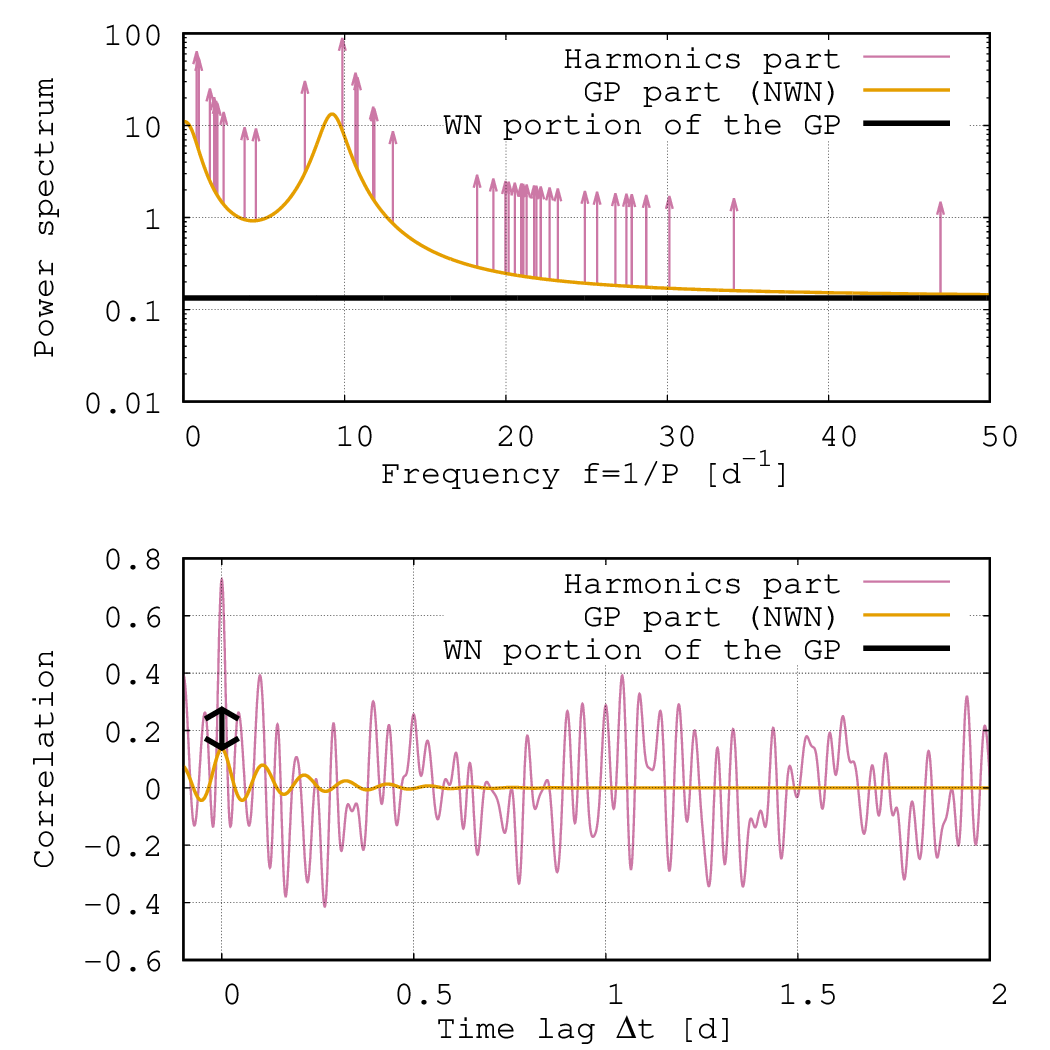}\\
\caption{Power spectrum and correlation function of the compound model ``harmonics+NWN''.}
\label{fig:pows}
\end{figure*}

In Fig.~\ref{fig:pows} we show the power spectrum of the compound model ``harmonics +
NWN'', as well as its correlation function. The power spectrum is normalized so that it can
be directly compared with the periodograms in Fig.~\ref{fig:prdgs}. The NWN has a
continuous power spectrum, but harmonics are expressed by delta functions there, so we plot
them as vertical arrows starting from the NWN graph. In the correlation function plots we
show separate graphs for the WN and NWN portions, and for the harmonics. Notice that
correlation function of a single sinusoid in~(\ref{mu}) is $\frac{A_k^2}{2} \cos 2\pi f_k
\Delta t$.

\section{Improving transit timing accuracy}
\label{sec_transits}
\subsection{Detrending and whitening the ITD data}
\label{sec_denoise}
Now let us consider how our out-of-transit photometric model of WASP-33 can be used to
detrend the transit lightcurves. The most direct and robust way to achieve this is to fit
the entire TESS dataset (OTD and ITD) with this model, augmented by the models of all the
transits. However, this way it would require too slow nonlinear fits, and besides we have a
goal in mind to employ our model with any other transit observations, so it is necessary to
develop an approach that avoids using full TESS data directly.

Our model contains two components: the deterministic one $\mu(t)$ (harmonics) and the GP
part (NWN) expressed by the correlation function $k(\Delta t)$. Regarding the deterministic
part, its uncertainty can be computed based on the covariance matrix of fitted parameters.
This uncertainty remains in the range $100$~ppm to $250$~ppm (for TESS$_{18}$ and
TESS$_{58}$, resp.) Relatively to the average TESS-reported uncertainty for these data,
$\sim 30000$~ppm, this appears below $1\%$. Therefore, possible uncertainty in $\mu(t)$ is
surely negligible in comparison with the best expected photometric accuracy. So we can
simply subtract our best fitting $\mu(t)$ from the ITD, as if it was a predefined function.
We will refer this part of the ITD processing as detrending for shortness. However, notice
that this differs from the detrending used in TESS data product (LC\_DETREND), even though
their practical effect may seem partly similar.

Regarding the NWN part of the model, its smallest $\tau_k$ from Table~\ref{tab:NWN} are
about $3$-$4$~h, which is similar to the exoplanet transit duration. This makes it
tentative to construct some kind of a GP prediction model (based on the OTD subset) and
then to subtract it from the ITD, just like $\mu(t)$. We undertook an attempt to compute an
OTD-conditioned GP predictive model with parameters taken from Table~\ref{tab:NWN}.
However, the practical predictability timescale appeared significantly smaller than
$\tau_k$, namely below $1$~h. Along the transit duration the variance of this prediction
remained mostly close to the total variance of the NWN (i.e. to the unconditional variance,
as if OTD were not used). Therefore, such a predictive model would not be very helpful, but
simultaneously this means that we would not loose significant information if we do not use
the OTD directly when processing the ITD. This enabled us to apply a simplified approach
with maximum-likelihood fitting of the ITD adopting a fixed covariance matrix (which is
defined by the NWN parameters from Table~\ref{tab:NWN}).

Importantly, this approach is equivalent to the whitening in which the data are passed
through a linear transform that renders them white. Formally, if $\mathbfss V = \Var
\bmath x$ is the covariance matrix of the data $\bmath x$, and $\bmath r$ is the vector
of their residuals then the modified (``whitened'') vector
\begin{equation}
\bmath r' = \mathbfss V^{-\frac{1}{2}} \bmath r
\end{equation}
appears such that
\begin{equation}
\Var \bmath r' = \mathbfss I, \quad |\bmath r'|^2 = \bmath r^{\rm T} \mathbfss V^{-1} \bmath r.
\end{equation}
The last quantity, $|\bmath r'|^2$, forms the only variable part of~(\ref{objf}) when
$\mathbfss V$ is fixed. Therefore, with our GP-fitting algorithm it is unnecessary to apply
whitening transform explicitly, but nonetheless we will refer this part of ITD processing
as whitening. This should be conceptually the same as the whitening used in TESS data
products (LC\_WHITE), but in practice the result largely depends on the adopted $\mathbfss
V$ which was different.

\begin{figure*}
\includegraphics[trim={0.45cm 0 0.55cm 0},clip,height=0.533,height=0.6125\linewidth]{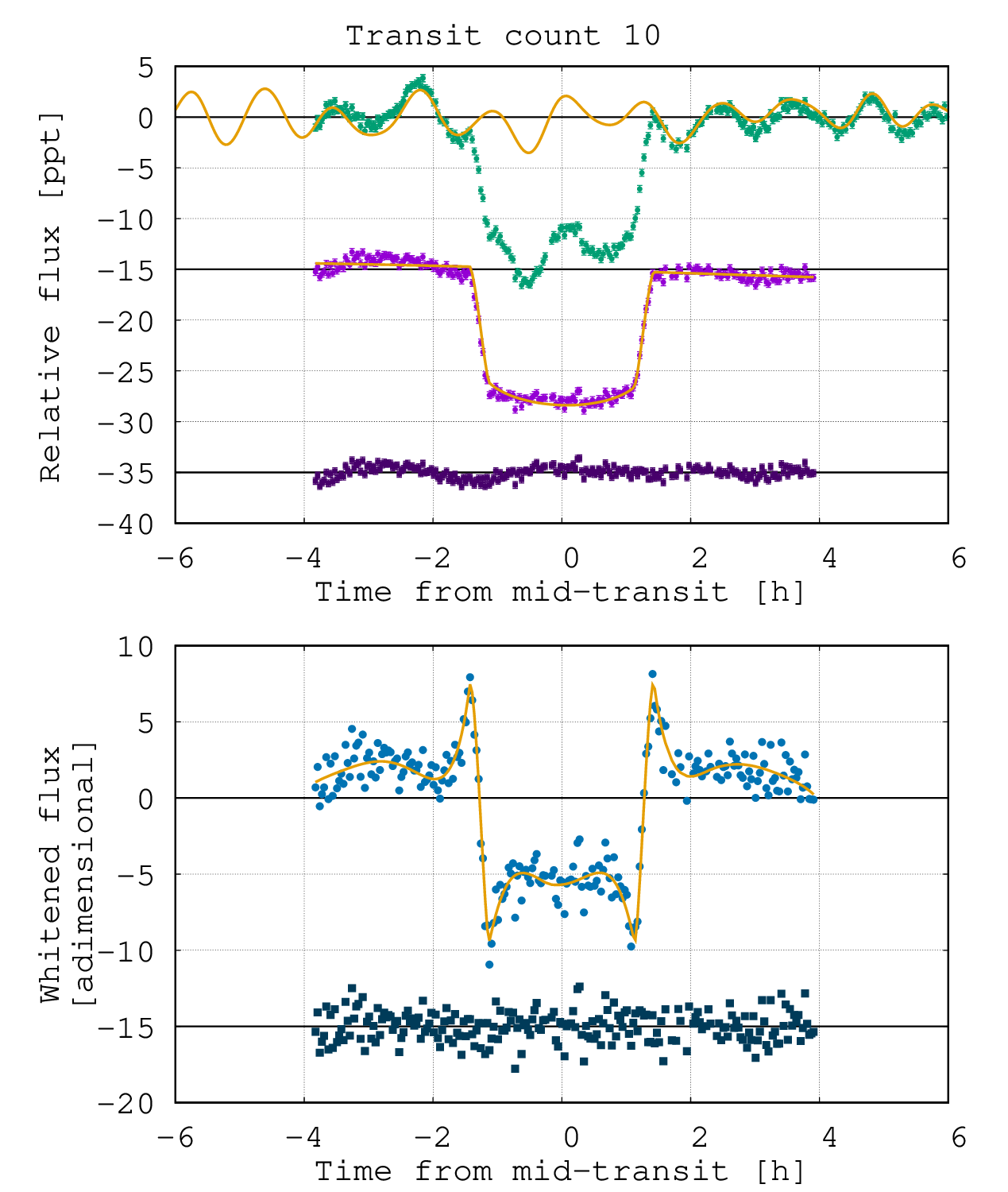}
\includegraphics[trim={2.09cm 0 0.55cm 0},clip,height=0.533,height=0.6125\linewidth]{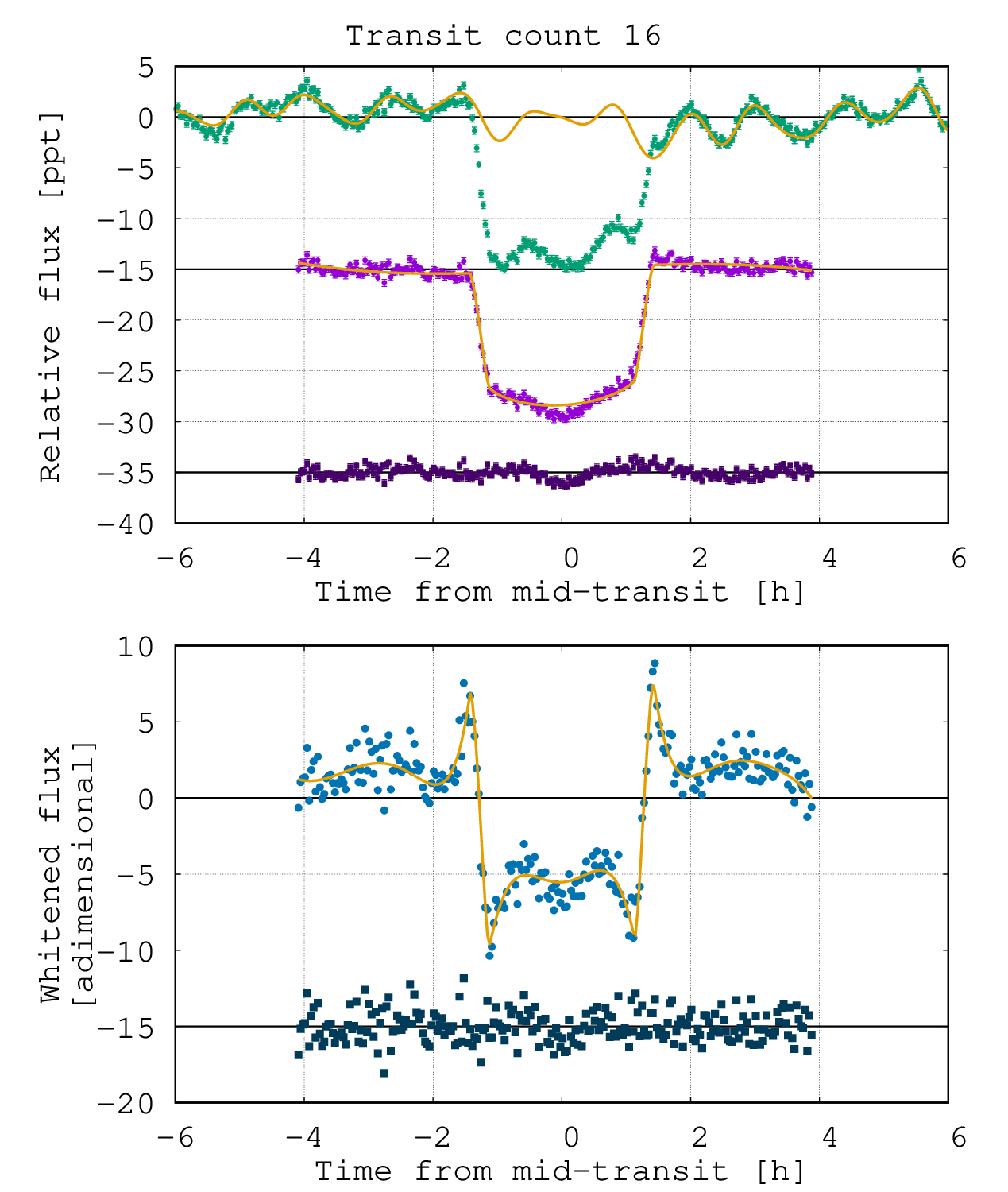}\\
\includegraphics[trim={0.45cm 0 0.55cm 0},clip,height=0.533,height=0.6125\linewidth]{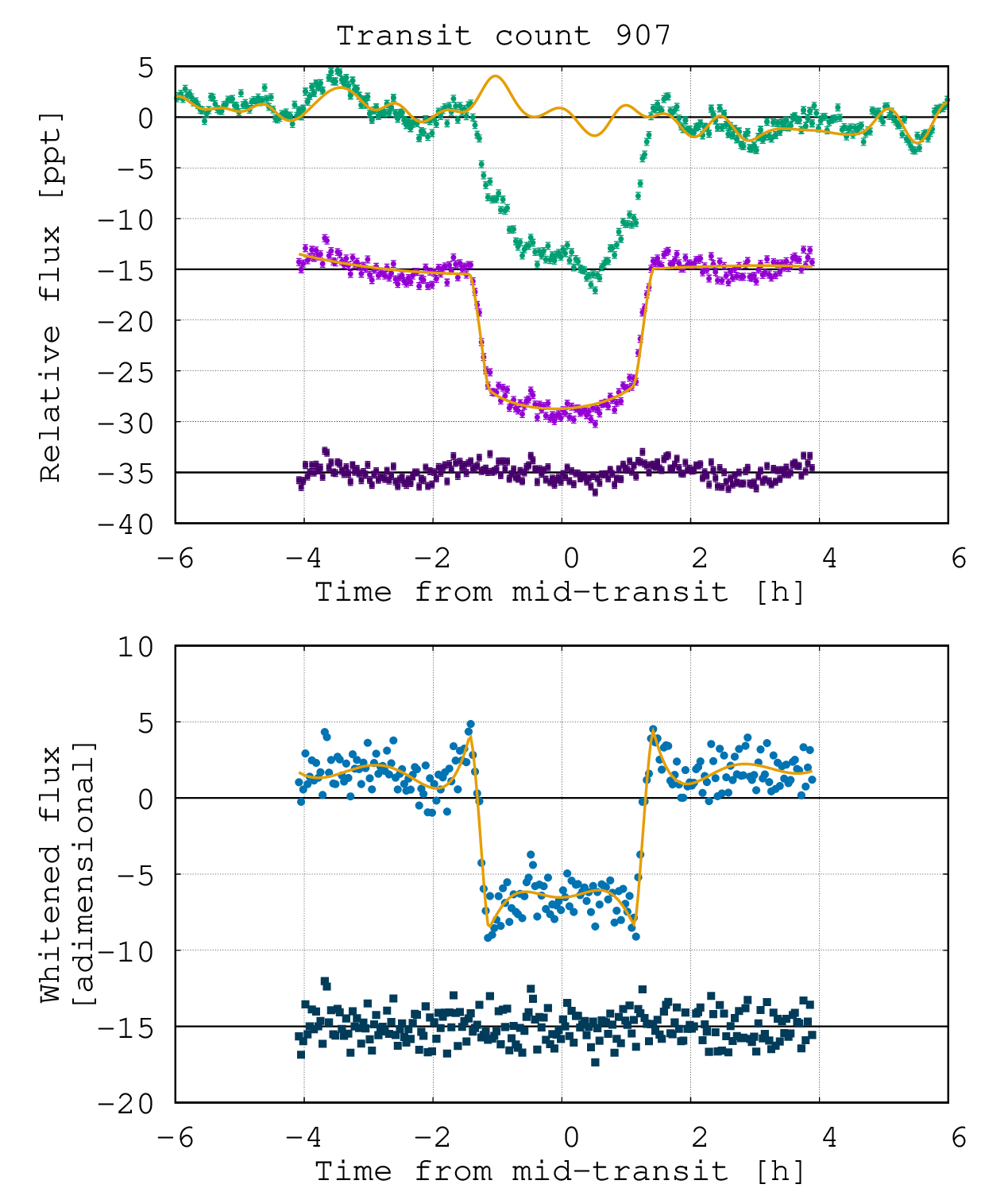}
\includegraphics[trim={2.09cm 0 0.55cm 0},clip,height=0.533,height=0.6125\linewidth]{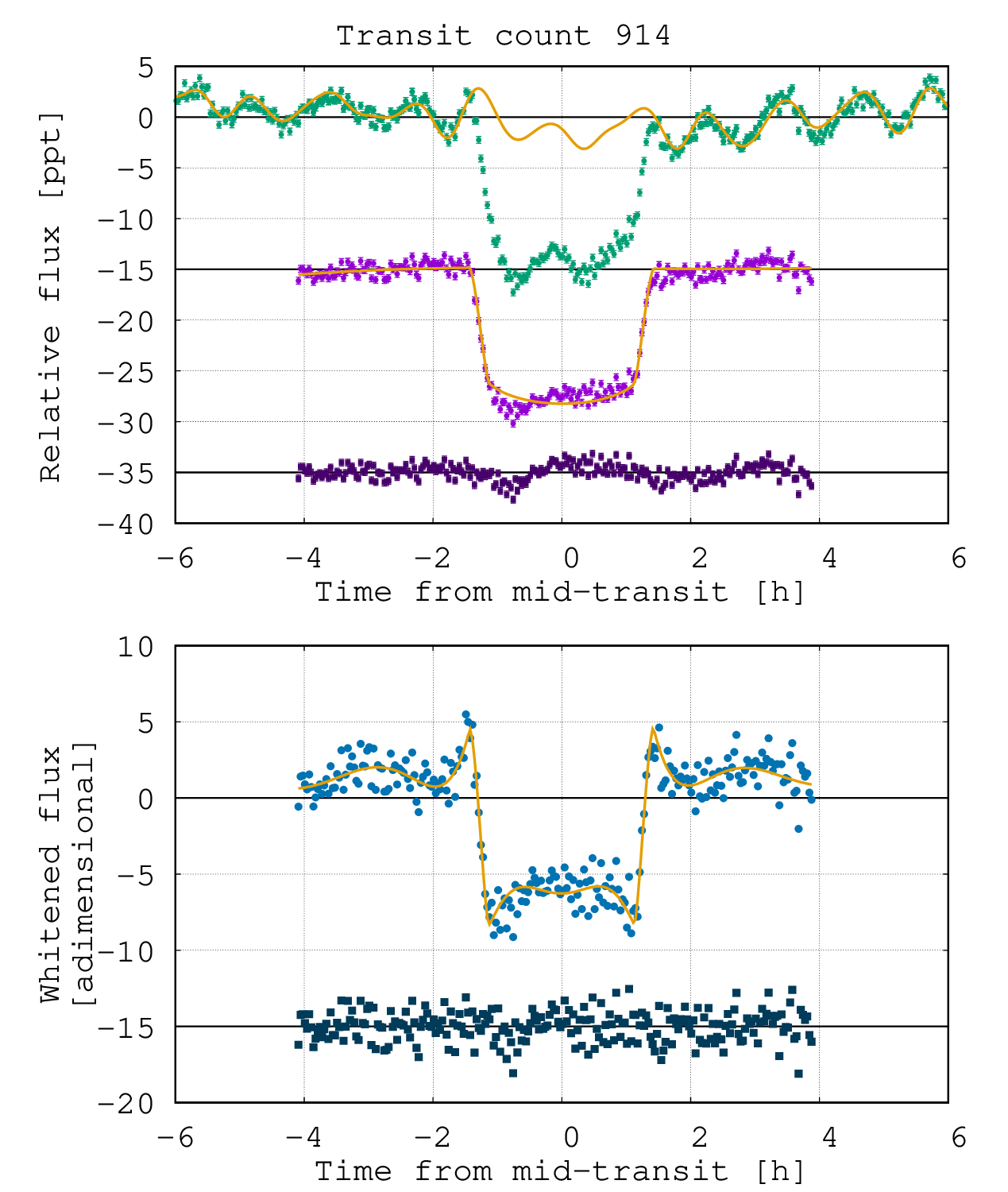}\\
\caption{Detrending and whitening for a few WASP-33\,\emph{b} transits. Notice that
whitening normalizes the data to unit variance, changing their scale. Since this variance
is always unit, the error bars for the whitened data are omitted for clarity. Models
residuals are plotted in the bottom of each graph (with a constant subtracted). Same plots
for all $38$ transits can be found in the online-only supplement.}
\label{fig:transits}
\end{figure*}

Summarising, our ITD processing consists of two cleaning steps, detrending and whitening.
In Fig.~\ref{fig:transits} we show how they work for a few sample transits. In these graphs
we plot the TESS data before processing (green points), the sum-of-harmonics $\mu$ (orange
curve), the detrended ITD (magenta points, shifted down) together with the transit model
(orange curve), the corresponding residuals (shifted down yet more), and detrended and
whitened ITD (blue points), also together with their model (which is also whitened) and
residuals (shifted down). One can see that detrending can remove a large portion of the
host star variability, however some residual nuisance variation still remains and should be
removed by whitening.

\subsection{Transit lightcurves fit}
We applied an NWN algorithm for a self-consistent fit of all our $38$ transits, which was
largely similar to the one presented in \citep{Baluev19}. The flux model involved a cubic
trend with fittable coefficients (individual per each transit), multiplied by the transit
lightcurve model by \citet{MandelAgol02} with quadratic limb darkening. We omitted the
outliers detection stage (since outliers were already removed in the DVT data), and adopted
fittable (rather than fixed) limb darkening coefficients. The NWN was modelled, whenever
appropriate, using the same GP technique and the same correlation models as described in
Sect.~\ref{sec_models}.

\begin{table*}
\caption{Best fitting WASP-33 system parameters.}
\label{tab:fit}
\begin{center}
\begin{tabular}{l@{}lr@{}ll@{}lr@{}llr@{}l}
\hline
\multicolumn{3}{c}{Planet}                              &      & \multicolumn{3}{c}{Star}                                 &         & \multicolumn{2}{c}{Limb darkening}\\
\hline
$P$              & [d]            & $1.219870527(70)$   &      & $M_\star^{\rm CC10}$ & [$M_\odot$]& $1.495(31)$          & $^{**}$ & $A$             & $0.212(34)$ & \\
$l$              & [$^\circ$]     & $86.412(10)$        & $^*$ & $R_\star$        & [$R_\odot$]    & $1.5101(68)$         &         & $B$             & $0.097(54)$ & \\
$i$              & [$^\circ$]     & $87.21(48)$         &      & $\rho_\star$     & [$\rho_\odot$] & $0.4342(59)$         &         & $A^{\rm C17}$   & $0.1883$    & $^{***}$\\
$r/R_\star$      &                & $0.11078(25)$       &$^{****}$&$M_\star^{\rm DR3}$&[$M_\odot$] & $1.54(4)$            &$^{***}$ & $B^{\rm C17}$   & $0.2873$    & $^{***}$\\
$r$              & [$R_{\rm Jup}$]& $1.665(10)$         &      & $R_\star^{\rm DR3}$ & [$R_\odot$] & $1.59(3)$            &$^{***}$ & $A^{\rm vE20}$   & $0.246(6)$  & $^{***}$\\
$a$              & [AU]           & $0.02555(35)$       &      &                  &                &                      &         & $B^{\rm vE20}$   & $0.252(6)$  & $^{***}$\\
\hline
\end{tabular}
\end{center}
\begin{minipage}{12cm}
$^{*\phantom{***}}$ Epoch $\rm BJD_{TDB} = 2459000$.\\
$^{**\phantom{**}}$ Taken from \citep{CCameron10b}.\\
$^{***\phantom{*}}$ Not used in the fit and given for comparison only (see text).\\
$^{****}$ This value is considerably affected by the third light correction, so its
uncertainty may be unreliable.
\end{minipage}
\end{table*}

Fitting the detrended and whitened ITD allowed us to obtain an improved set of WASP-33
system parameters, given in Table~\ref{tab:fit}. This model assumes that transit timings
strictly follow a linear ephemeris (no TTV). The fit included the stellar mass estimate by
\citet{CCameron10b} which was based on their spectrum model, line-profile tomography, and
ground-based transit data constraining stellar density. This stellar mass was accounted for
through a penalty function added to the log-likelihood \citep[see][]{Galazutdinov23}. We
ignored the accompanying stellar radius value since we need to know the correlation between
$M_\star$ and $R_\star$ to correctly include them both. Our estimate for the stellar
density appeared smaller by $\sim 2.5\sigma$ than that by \citet{CCameron10b}.

Independent estimations of WASP-33 mass and radius are available in GAIA DR3 FLAME data
table \citep{Gaia23,GaiaAP}; they were added to Table~\ref{tab:fit} for comparison.
Although these estimations use more accurate parallax, we did not use them in our fit,
because contrary to \citet{CCameron10b} they (i) are based on a rather low-resolution
spectrum, (ii) did not include constraints from any transit observations of
WASP-33\,\emph{b}. This resulted in a considerably larger uncertainties in GAIA $M_\star$,
$R_\star$. We were also bothered by very low metallicity of $-0.56$ given in the same GAIA
table and looking inconsistent with e.g. \citet{CCameron10b}. Yet another independent
$M_\star$, $R_\star$ estimations provided by \citet{vonEssen20} have even larger
uncertainties than GAIA ones.

In addition to the best fitting estimates of limb darkening coefficients, we also give in
Table~\ref{tab:fit} their reference values (C17) computed from the tables by
\citet{Claret17} using $T_{\rm eff}$, $\log g$, and [Fe/H] by \citet{CCameron10b} and
(vE20) computed by \citet{vonEssen20}. These reference sets of estimates both differ from
the fitted one, as well as from each other. This may indicate some issues regarding their
reliability for not yet clear reasons.

\subsection{Transit timings}
\begin{figure*}
\includegraphics[width=\linewidth]{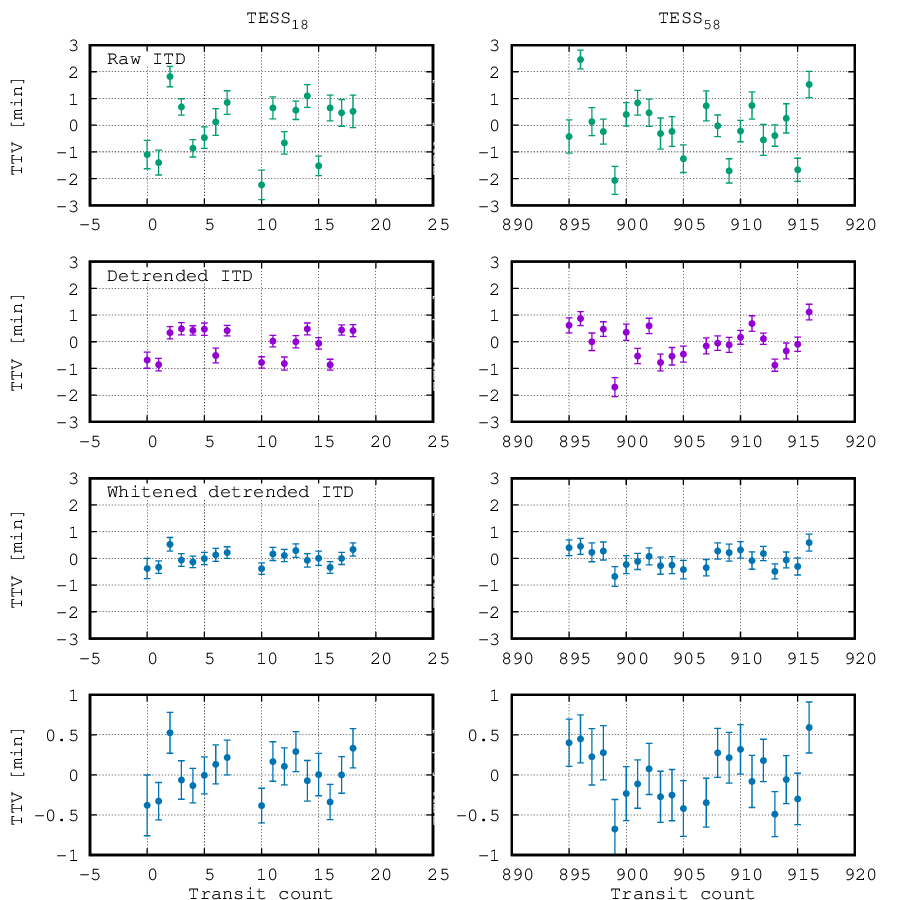}
\caption{WASP-33 transit timing residuals before and after ITD cleaning. In the bottom pair
of graphs the whitened detrended sets were replotted using larger ordinate scale.}
\label{fig:ttv}
\end{figure*}

On the final step we fitted the ITD once again, but assuming fittable transit timings. This
was done for all three types of the ITD: raw, detrended and whitened detrended. The effect
of our cleaning method is demonstrated in Fig.~\ref{fig:ttv}, where we plot the TTV
residuals derived from the initial (raw) ITD, detrended ITD and detrended and whitened ITD.
One can see that the TTV scatter is remarkably reduced after each step. The initial TTV
residuals r.m.s. was $63$~s. After detrending it became $36$~s, while after the whitening
it finally dropped to $18$~s. In total we have a rather dramatic r.m.s. improvement by a
factor of $3.5$.

\begin{table*}
\caption{WASP-33\,\emph{b} transit timings derived from the TESS ITD.}
\label{tab:ttv}
\footnotesize
\begin{center}
\begin{tabular}{ccccccc}
\hline
Count &Transit midtime&Uncertainty& & Count &Transit midtime&Uncertainty\\
      &${\rm BJD}_{\rm TDB}-2450000$&[s] & & &${\rm BJD}_{\rm TDB}-2450000$&[s]\\
\hline
$0 $ & $8791.414035$ & $22.7$ & & $895$ & $9883.198699$ & $17.7$ \\
$1 $ & $8792.633942$ & $14.0$ & & $896$ & $9884.418604$ & $18.0$ \\
$2 $ & $8793.854405$ & $15.3$ & & $897$ & $9885.638319$ & $21.2$ \\
$3 $ & $8795.073867$ & $14.4$ & & $898$ & $9886.858225$ & $20.3$ \\
$4 $ & $8796.293688$ & $13.0$ & & $899$ & $9888.077433$ & $22.2$ \\
$5 $ & $8797.513648$ & $13.8$ & & $900$ & $9889.297612$ & $20.2$ \\
$6 $ & $8798.733613$ & $14.7$ & & $901$ & $9890.517565$ & $18.1$ \\
$7 $ & $8799.953544$ & $13.0$ & & $902$ & $9891.737567$ & $19.2$ \\
$10$ & $8803.612739$ & $13.0$ & & $903$ & $9892.957195$ & $19.2$ \\
$11$ & $8804.832991$ & $14.8$ & & $904$ & $9894.177081$ & $19.2$ \\
$12$ & $8806.052820$ & $13.8$ & & $905$ & $9895.396834$ & $20.9$ \\
$13$ & $8807.272818$ & $14.9$ & & $907$ & $9897.836626$ & $18.3$ \\
$14$ & $8808.492436$ & $15.2$ & & $908$ & $9899.056929$ & $18.3$ \\
$15$ & $8809.712360$ & $15.9$ & & $909$ & $9900.276757$ & $19.0$ \\
$16$ & $8810.931993$ & $13.2$ & & $910$ & $9901.496700$ & $18.5$ \\
$17$ & $8812.152098$ & $13.7$ & & $911$ & $9902.716293$ & $19.5$ \\
$18$ & $8813.372200$ & $14.7$ & & $912$ & $9903.936345$ & $15.9$ \\
     &               &        & & $913$ & $9905.155751$ & $16.8$ \\
     &               &        & & $914$ & $9906.375920$ & $18.0$ \\
     &               &        & & $915$ & $9907.595623$ & $19.3$ \\
     &               &        & & $916$ & $9908.816113$ & $19.1$ \\
\hline
\end{tabular}\\
\emph{NB} Machine-friendly version of this table, with uncertainties given in days, can be
found in the online-only supplement.
\end{center}
\end{table*}

The last best-accuracy set of transit timings is given in Table~\ref{tab:ttv}, while in
Table~\ref{tab:ttvdat} we provide some statistics regarding the fitted timings for all
three types of the ITD, and separately for TESS sectors. One can see that only the full
cleaning makes our fit-derived timing uncertainties to agree with their actual SSD. In
other cases the SSD appears larger than formal uncertainties, indicating inaccuracy of the
noise model. We also note that TESS data reveal remarkable homogeneity and stability over a
single sector, which is indicated by very small values of uncertainty SSD (last column in
Table~\ref{tab:ttvdat}), compared to the uncertainty mean value. There is some difference
in the average uncertainty between sectors, but it corresponds to just about $1.6\sigma$,
which is not enough for reliable conclusions.

\begin{table*}
\caption{Statistics of the resulting transit timings and their uncertainties.}
\label{tab:ttvdat}
\footnotesize
\begin{center}
\begin{tabular}{lcccccc}
\hline
Data type  & Sector & Mean timing residual [s] & Residuals SSD$^*$ [s] & Mean timing uncertainty [s] & Uncertainties SSD$^*$ [s] \\
\hline
Raw        & 18     & $-3.0$   & $65.7$   & $26.4$        & $5.0$    \\
           & 58     & $-4.4$   & $64.3$   & $29.5$        & $4.3$    \\
Detrended  & 18     & $-3.8$   & $33.4$   & $13.5$        & $1.8$    \\
           & 58     & $-1.9$   & $39.8$   & $17.2$        & $2.0$    \\
Detrended +& 18     & $+0.2$   & $15.8$   & $14.7$        & $2.2$    \\
\ whitened & 58     & $-0.7$   & $20.6$   & $19.0$        & $1.5$    \\
\hline
\end{tabular}\\
$^*$ Sample Standard Deviation.
\end{center}
\end{table*}

We also investigated timing data from Table~\ref{tab:ttv} for possible variations, but with
a null result: we did not find any significant quadratic term in the TTV trend and we did
not find any significant periodic TTV.

\section{Discussion}
The approach based on the explicit modelling of WASP-33 variability appears extremely
successful in improving its exoplanet transit timings accuracy. In total, we obtained a
dramatic TTV r.m.s. improvement by a factor of $3.5$. It is important that we must take
into account the deterministic as well as non-deterministic (NWN) parts of this
variability. Both of them contribute a similar amount in the total TTV error budget.

Our results still cannot be directly applied to observations from other sites than TESS,
because the model parameters somewhat change on the timescale of a few years. Because of
that we cannot use TESS-derived model with data obtained e.g. years prior to TESS. However,
the frequencies of the harmonics and of the QP part of the NWN look stable, so it may be
possible to apply our model to ground-based WASP-33 photometry assuming that only these
frequencies are fixed and to refit harmonics amplitudes and phases, as well as NWN
variances. But possible ways to reprocess old data using TESS-based models need more
investigation.

Our results regarding the deterministic part of the WASP-33 variability are generally
similar to \citet{vonEssen20}. We found $43$ harmonics in total, although only $30$ of them
are surely reliable because they were revealed in the both datasets, TESS$_{18}$ and
TESS$_{58}$. \citet{vonEssen20} detected $29$ harmonics in TESS$_{18}$, which is nearly the
same number. However, there are a few individual differences. Based on our analysis we
cannot confirm that their harmonics F21, F25, and F29 are statistically significant, even
in a single TESS sector. On contrary, our harmonics H22, H32, H37, and H40 were revealed
both in TESS$_{18}$ and TESS$_{58}$, but left undetected by \citet{vonEssen20}. Among our
harmonics that were revealed in TESS$_{18}$, but not in TESS$_{58}$, only one (H3 = F28)
appeared in \citet{vonEssen20}. Among harmonics that were detected in TESS$_{58}$, but not
in TESS$_{18}$, none appeared in \citet{vonEssen20}. This may render our $13$
``single-sector'' harmonics less reliable, although this does not necessarily mean that all
of them, or their majority, are false positives. Based on our FAP threshold of $5\%$, the
expected number of false positives after $36$ or $37$ applications of the significance test
should be just one or two (for each TESS sector). And these false positives are, most
probably, among the ``single-sector'' harmonics.

Regarding the secondary eclipses of WASP-33\,\emph{b}, detected by \citet{vonEssen20}, we
cut them away from our OTD, so they did not affect our analysis and we did not consider
them here. But two other planet-related effects reported by \citep{vonEssen20}, namely the
orbital phase variation and the ellipsoidal variation, could affect our results, so we
included them in the model $\mu$ from the beginning of our analysis. For the orbital phase
variation our final amplitude estimate was $A_{\rm pv}=(110\pm 100)$~ppm and $(32\pm
44)$~ppm (resp. in TESS$_{18}$ and TESS$_{58}$), while \citet{vonEssen20} gave $A_{\rm
pv}=(100.4 \pm 13.1)$~ppm.

One can see that our uncertainty for $A_{\rm pv}$ is much larger and this requires
investigation. We considered a number of possible explanations for such difference. First,
our analysis used a reduced phase coverage because we cut away transits and secondary
eclipses. In total, an $8$ hours portion of each orbital period was removed, which is only
$27$ per cent of the full sector~18 data. This should leave about $90$ per cent of the
$A_{\rm pv}$ variation, and clearly could not increase its uncertainty by a factor of $7$.
Secondly, we tried to seek and cut away potentially unreliable parts of the data. In
particular in the beginning of the TESS$_{18}$ lightcurve the spacecraft passed through the
Earth shadow that changed its temperature mode. Removing a small initial portion before the
passage resulted in $A_{\rm pv} = (120\pm 100)$~ppm. Finally we considered the raw SAP
data. We applied a simple detrending technique that was discussed in
Sect.~\ref{sec_select}, again removed the segment before the Earth shadow passage, and then
refitted our compound model ``NWN+harmonics'' using these data. We obtained $A_{\rm pv} =
(130\pm 80)$~ppm after that. In fact we found the only way to reproduce $\sigma_{A_{\rm
pv}} \sim 13$~ppm, which was to remove the NWN treatment entirely. By fitting the data as
if the noise was purely white, we had $A_{\rm pv} = (130\pm 14)$~ppm.

This looked as if the issue is related to how the NWN was processed, and therefore we tried
to reproduce \citet{vonEssen20} methods at least partly. We applied the $\beta$ method by
\citet{Winn08} to our OTD, which resulted in the maximum $\beta\simeq 11$ (achieved
for the bin width about $1$~d). Given this value, we should scale our WN-derived
uncertainty up from $\pm 14$~ppm to $\pm 150$~ppm. Therefore, if we used $\beta$ method in
our study instead of GP fitting, we would obtain even larger uncertainty for $A_{\rm pv}$.

Detailed investigation of other parameters fitted by \citet{vonEssen20} revealed a yet
another cautioning matter. Literally all amplitudes of the pulsation harmonics have exactly
the same uncertainty of $\pm 10$~ppm (see their Table~2). However, such condition is
impossible for correlated noise, because its magnitude varies over the frequency axis. In
case of WASP-33 its power spectrum is peaked at low frequencies, so harmonics located in
this range must have larger uncertainties than high-frequency ones, which are practically
unaffected by the NWN. Such a behaviour is obvious in Table~\ref{tab:harmonics}, where the
uncertainties of $A_k$ increase by a factor of $\sim 5$ from top line to bottom.

The fact the all $\sigma_{A_k}$ appeared the same in \citep{vonEssen20} might have two
alternative explanations. The first one is due to the $\beta$ method which in fact does not
offer a genuine NWN treatment: it simply scales all uncertainties up by the most
conservative multiplier. In this case uncertainties of low-frequency harmonics would be
more adequate, but those at high frequencies would be considerably overvalued. The opposite
explanation is that high-frequency harmonics are adequate, while low-frequency ones are
considerably less constrained than it follows from the stated uncertainties.

To determine which situation takes place here, let us assess the uncertainty of harmonics
amplitudes, assuming that noise is white. There is a well-known approximation of this
uncertainty \citep[eqs~6.17-6.19]{Vityazev94}\footnote{Formally, this approximation
requires that the spectral window is small at all frequencies $f_i\pm f_j$, where $f_i$
being the $i$th harmonic frequency (this set includes $2f_i$, but excludes zero). The TESS
time series are practically even, so the spectral window is always small in the domain of
interest.}:
\begin{equation}
\sigma_A \simeq \sigma_{\rm noise} \sqrt{\frac{2}{N}}.
\end{equation}
From Fig.~10 by \citet{vonEssen20} the value of $\sigma_{\rm noise}$ follows to be about
$700$~ppm, while $N$ was $14000$. Therefore, for a WN fit these data would imply
$\sigma_A\simeq 8.4$~ppm. If NWN is being corrected with the $\beta$ method, a WN-based
$\sigma_A$ should be multiplied by the factor $\beta=3.545$, as adopted by
\citet{vonEssen20}, resulting in $\sigma_A\sim 30$~ppm. This is basically a minimum
realistic value for an analysis that honoured NWN. However, in actuality they give
$\sigma_A=10$~ppm, close to the pure WN case.

Based on the investigation presented above we believe that by some hidden mistake, or
because of a software bug, \citet{vonEssen20} uncertainties for the harmonics $F_k$ refer
to the WN treatment instead of the NWN one. Most probably, the same holds true for the
putative phase variation. This harmonic should be correlated with the transit and secondary
eclipse parameters, and hence $\sigma_{A_{\rm pv}}$ may be even larger than $\sigma_{A_k}$.
According to \citet{vonEssen20}, $\sigma_{A_{\rm pv}}=13.1$~ppm is a bit larger than
$\sigma_{A_k}=10$~ppm indeed, but not enough to be realistic for an NWN fit ($\sim 30$~ppm
at least). If we assume that $\pm 13.1$~ppm refered actually to a WN model, the NWN
uncertainty would be about $\pm 50$~ppm. We also believe that constant flux term $c_0$ has
an unrealistically small uncertainty too, while other fit parameters are less obvious.

From our GP analysis we cannot confirm that there is a statistically significant evidence
in favour of orbital phase variation in TESS data~--- not only because of a formally larger
uncertainty we obtained, but also because in the second TESS sector this estimation
appeared to differ from the first one.

For the ellipsoidal variation we obtained the amplitude of $(65\pm 41)$~ppm and $(97\pm
31)$~ppm, for the two TESS sectors. \citet{vonEssen20} gave $27.4$~ppm in the text,
although we think this could be a typo, because from their Fig.~9 this amplitude follows to
be $\sim 40$~ppm. The latter value is within at most two-sigma from our estimations, but
still it is comparable to the uncertainty and cannot be reliably confirmed observationally.
However, contrary to the phase variation, this amplitude is theoretically predicted and
represents something that required a reduction, rather than observational estimation or
confirmation.

\section*{Acknowledgements}
We are grateful to the anonymous reviewers for their useful and constructive suggestions
that helped to improve the manuscript.

\section*{Data availability}
The data underlying this article are available in the MAST (Mikulski Archive for Space
Telescopes), at \url{https://archive.stsci.edu/}, in the article and in its online
supplementary material.

\bibliographystyle{mnras}
\bibliography{WASP33}

\bsp

\label{lastpage}

\end{document}